\documentclass[iop,apj,tighten]{emulateapj}
\usepackage{mathtools, apjfonts, multirow, soul}
\usepackage[usenames,dvipsnames]{color}

\def\star{PTFO~8-8695}
\def\planet{{\star}b}

\def\um{\mu \mathrm{m}}

\def\asc{\lambda}
\def\obliq{\psi}
\def\spinorbit{\varphi}
\def\Lspin{L_{*}}
\def\Lorbit{L_{p}}
\def\Ltotal{L_{\mathrm{total}}}
\def\Lspinvec{{\vec{L}_*}}
\def\Lorbitvec{{\vec{L}_p}}
\def\Ltotalvec{{\vec{L}_\mathrm{total}}}
\def\LspinvecN{{\vec{L}_{*0}}}
\def\LorbitvecN{{\vec{L}_{p0}}}
\def\LtotalvecN{{\vec{L}_\mathrm{total0}}}

\def\transitfitter{{\tt transitfitter}}

\shorttitle{Measurement of Nodal Precession of \planet~from Gravity Darkening}
\shortauthors{Barnes \emph{et al.}}

\begin{document}

\title{Measurement of Spin-Orbit Misalignment and Nodal Precession for the Planet Around\\
Pre-Main-Sequence Star \star~from Gravity Darkening}

\author{\scshape Jason W. Barnes\altaffilmark{1}, Julian C. van
Eyken\altaffilmark{2}, Brian K. Jackson\altaffilmark{3}, David R.
Ciardi\altaffilmark{4}, Jonathan J. Fortney\altaffilmark{5}} 

\altaffiltext{1}{Department of Physics, University of Idaho, Moscow, ID 83844-0903,
USA:  ResearcherID:  B-1284-2009, {\tt jwbarnes@uidaho.edu}}
\altaffiltext{2}{Department of Physics, University of California Santa Barbara, 
Santa Barbara, CA, 93106-9530, USA}
\altaffiltext{3}{Carnegie Institution of Washington, DTM, 5241 Broad Branch Road, NW 
Washington, DC 20015-1305}
\altaffiltext{4}{NASA Exoplanet Science Institute, Caltech M/S 100-22, Pasadena, CA 91125, USA}
\altaffiltext{5}{Department of Astronomy, University of California Santa Cruz, 
Santa Cruz, CA, 95064, USA}

\begin{abstract} 

\planet~represents the first transiting exoplanet candidate orbiting a
pre-main-sequence star \citep{2012ApJ...755...42V}.  We find that the unusual
lightcurve shapes of \star~can be explained by transits of a planet across an
oblate, gravity-darkened stellar disk.  We develop a theoretical framework for
understanding precession of a planetary orbit's ascending node for the case when
the stellar rotational angular momentum and the planetary orbital angular
momentum are comparable in magnitude.  We then implement those ideas to
simultaneously and self-consistently fit two separate lightcurves observed in
2009 December and 2010 December.  Our two self-consistent fits yield
$M_p=3.0~\mathrm{M_{Jup}}$ and $M_p=3.6~\mathrm{M_{Jup}}$ for assumed stellar
masses of $M_*=0.34~\mathrm{M_\odot}$ and $M_*=0.44~\mathrm{M_\odot}$
respectively.  The two fits have precession periods of 293~days and 581~days. 
These mass determinations (consistent with previous upper limits) along with the
strength of the gravity-darkened precessing model together validate \planet~as
just the second Hot Jupiter known to orbit an M-dwarf.  Our fits show a high
degree of spin-orbit misalignment in the \star~system:  $69^\circ\pm2^\circ$ or
$73.1^\circ\pm0.5^\circ$, in the two cases.  The large misalignment is
consistent with the hypothesis that planets become Hot Jupiters with random
orbital plane alignments early in a system's lifetime.  We predict that as a
result of the highly misaligned, precessing system, the transits should
disappear for months at a time over the course of the system's precession
period.  The precessing, gravity-darkened model also predicts other observable
effects:  changing orbit inclination that could be detected by radial velocity
observations, changing stellar inclination that would manifest as varying $v\sin
i$, changing projected spin-orbit alignment that could be seen by the
Rossiter-McLaughlin effect, changing transit shapes over the course of the
precession, and differing lightcurves as a function of wavelength.  Our measured
planet radii of $1.64~\mathrm{R_{Jup}}$ and $1.68~\mathrm{R_{Jup}}$ in each case
are consistent with a young, hydrogen-dominated planet that results from a
`hot-start' formation mechanism.  

\end{abstract}

 
\section{INTRODUCTION}

The solar system planets all orbit in planes within $7.5^\circ$ of the Sun's
equator.  However, the orbits of 34\% of measured extrasolar planets show
statistically significant inclinations (greater than two standard deviations)
with respect to their star's equator \citep{2012ApJ...757...18A}.  We call this
situation spin-orbit misalignment.  

These numerous misaligned systems challenge our understanding of planet
formation and evolution.  The misaligned systems must not have formed in the
same way as our solar system planets (presuming that the Sun's low inclination
to the protoplanetary disk was not coincidental).  Because nearly all of the
extrasolar planets whose stellar spin-planetary orbit alignments have been
measured are Hot Jupiters, the mechanism for producing misalignment may relate
to the origins of Hot Jupiters.

\citet{2010ApJ...718L.145W} noted that Hot Jupiters around higher mass stars are
preferentially misaligned as compared to those around solar-type stars. 
\citet{2010ApJ...718L.145W} suggested that tidal realignment, which occurs
faster in low-mass stars with convective envelopes, could explain the spin-orbit
alignment dependence on stellar mass.  More recently \citet{2012ApJ...757...18A}
used spin-orbit alignment measurements as a function of stellar properties to
confirm that the observed distribution of alignments is consistent with tidal
realignment of initially random Hot Jupiter orbit orientations.

The discovery of a transiting planet candidate around a pre-main-sequence (PMS)
low-mass star by \citet{2012ApJ...755...42V} can shed
light on the origin of Hot Jupiter misalignment.  The host star, \star, is an M
dwarf with $0.44~\mathrm{M_\odot}$ or $0.34~\mathrm{M_\odot}$ (depending on the
model) and an effective temperature of just 3470 K \citep{2005AJ....129..907B}. 
Its age is 2.63-2.68~Myr \citep{2005AJ....129..907B}.  Hence, a determination of
the spin-orbit alignment angle for the putative planet, \planet, would represent
such a measurement for the smallest, coldest, and youngest planet-hosting star.

The transit lightcurve for \planet~observed by \citet{2012ApJ...755...42V} shows
an unusual shape that changes between observations acquired a year apart.  The
transit depth is greater and its duration shorter in the second
observation.  \citet{2009ApJ...705..683B} showed that unusual, asymmetric shapes
can result from transits across rapidly-rotating stars.  The lower effective
gravity at these stars' equators results in cooler effective temperatures there
relative to the stars' poles \citep{1924MNRAS..84..665V}, which can lead to
transit shapes distinct from those for stars with only limb darkening.
  
\begin{figure}[htbp]
\epsscale{1.1}
\plotone{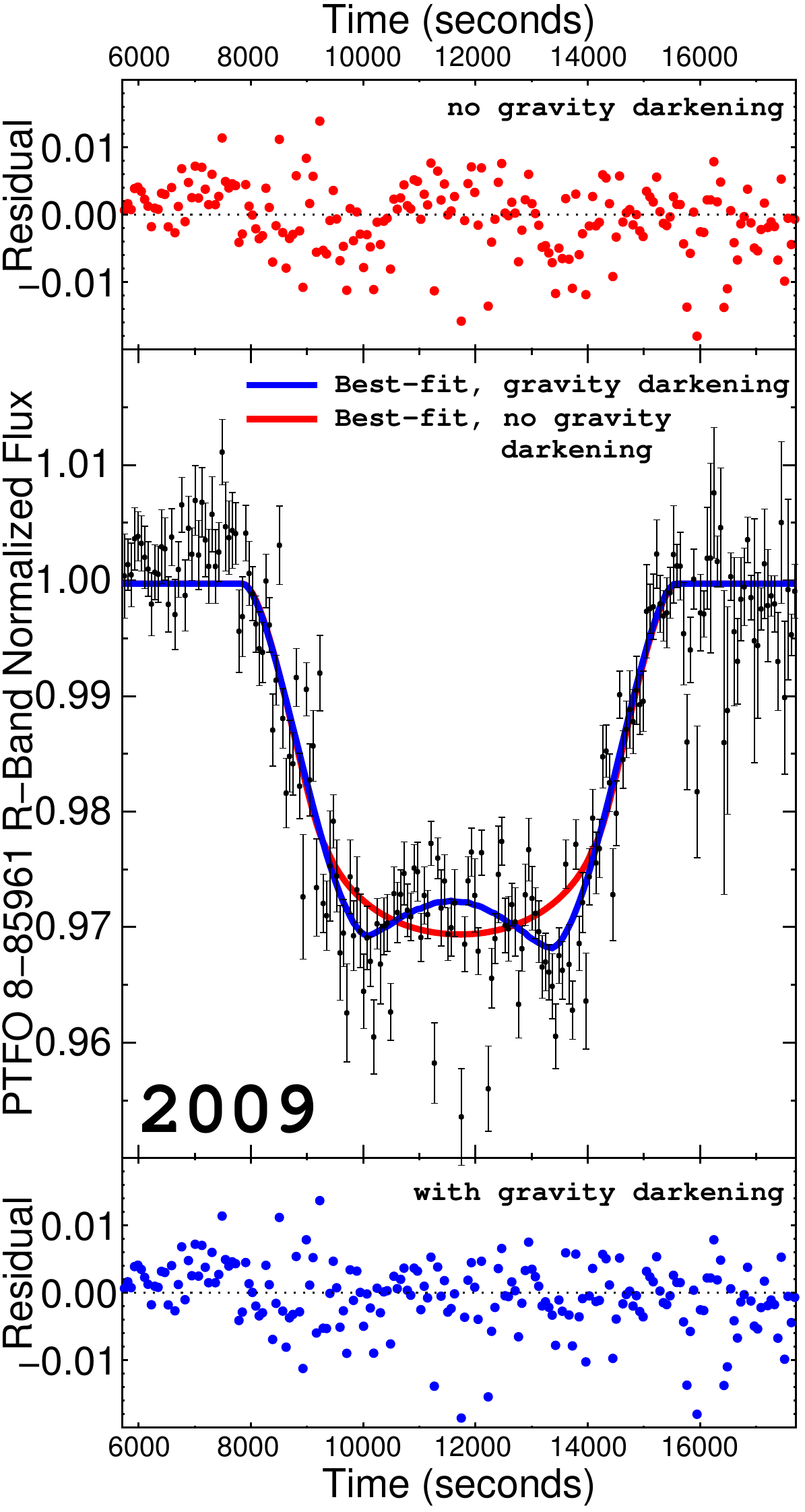}
\caption{\footnotesize Photometry and fits for the phase-folded 2009 \planet~lightcurve.  We
plot the
data and fits themselves in the center with the best-fit conventional,
no-gravity-darkening model (\emph{i.e.} with a spherical star) in red and the 
gravity-darkened model in blue.  The residuals from both fits are shown at top
(spherical) and bottom (gravity-darkened).  The gravity-darkened model does a
reasonable job of reproducing the convexity at the bottom of the lightcurve. 
\label{figure:2009_fit}}
\end{figure}

Precession of the ascending node of \planet's orbit and/or the rotation pole for
\star~could allow gravity darkening to then explain the changes in transit depth
and duration changes from 2009 to 2010.  Nodal precession has been seen in one
other misaligned planetary system so far:  KOI-13 
\citep{2012MNRAS.421L.122S,2011ApJS..197...10B,2011ApJ...736L...4S}.  Gravity
darkening has been successfully invoked previously to explain varying shapes of
the lightcurves of eclipsing binary stars as well \citep{DIHerculis}.

In this paper, we investigate whether precession of the planet-star system
combined with stellar gravity darkening can explain the unusual lightcurves for
\planet~seen by \citet{2012ApJ...755...42V} and as we describe in Section
\ref{section:observations}.  We start off by describing the
\citet{2012ApJ...755...42V} observations in Section \ref{section:observations}. 
In Section \ref{section:individualfits} we examine the 2009 and 2010 lightcurves
separately in the context of gravity darkening.  Then in 
\ref{section:precession} and \ref{section:model} we develop the theory and a
numerical model for orbital precession of Hot Jupiters.  We show in
Section \ref{section:extrapolation} how that precession would affect the
individual fits.  And in Section \ref{section:jointfit}, we discuss a
self-consistent joint fit that can model the transit observations from both 2009
and 2010 simultaneously.  We discuss the implications that the joint
fit has on future observations in Section \ref{section:projection} before a
discussion and conclusion in Section \ref{section:discussion}.

\begin{figure}[htbp]
\epsscale{1.1}
\plotone{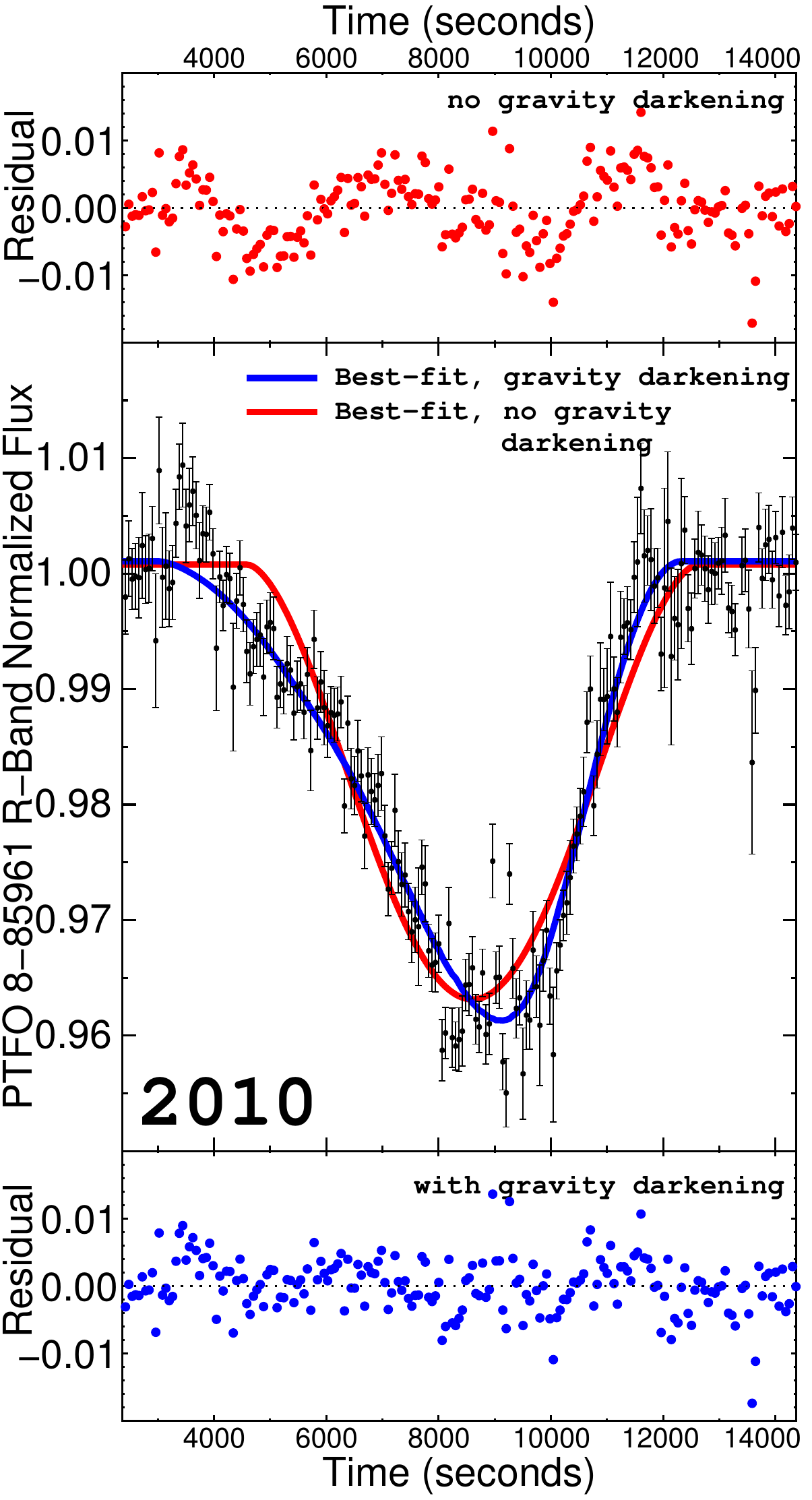}
\caption{\footnotesize Photometry and fits for the phase-folded 2010 \planet~lightcurve.  We
plot the
data and fits themselves in the center, with the best-fit conventional,
no-gravity-darkening model (\emph{i.e.} with a spherical star) in red and the 
gravity-darkened model in blue.  The residuals from both fits are shown at top
(spherical) and bottom (gravity-darkened).  Without gravity darkening the fit is
poor.  With gravity darkening, the fit is able to tune aggressively to match the
transit duration, sharp transit bottom, and long ingress tail simultaneously. 
\label{figure:2010_fit}}
\end{figure}

\section{OBSERVATIONS} \label{section:observations}

We introduce no new observations in this paper.  Instead we reanalyze photometry
of \star~as published in the \planet~discovery paper by
\citet{2012ApJ...755...42V}.  As part of the the Palomar Transient Factory Orion
(PTFO) campaign, \citet{2012ApJ...755...42V} acquired R-band relative photometry
of \star~during all or portions of 11 separate transits of \planet~between 2009
December 3 and 2010 January 14 (which we refer to as the ``2009" observations),
and 6 separate transits between 2010 December 8 and 2010 December 14.  

As a T-Tauri star, \star~shows significant amounts of stellar variability owing
to starspots and activity.  These spots manifest as red noise in the resulting
lightcurve.  The stellar rotational modulation induced by starspots
occurs on a much longer timescale than the transit, and thus we remove it using
a spline fit to the out-of-transit points.  However, variations in the
lightcurve that might result from the planet passing over individual starspots
or starspot clusters likely remain.  Such starspot crossings have been seen for
other transiting planets \citep[\emph{e.g.}][]{2007A&A...476.1347P}, and have in
some cases been used to measure spin-orbit alignment via the pattern of starspot
crossings on successive transits
\citep[\emph{e.g.}][]{2011ApJS..197...14D,2011ApJ...740L..10N}.  Because the
typical lifetime for sunspots is days to a couple of weeks, any starspot effects
on the transit lightcurve shape in the \star~system should decohere
on that timescale.

Therefore to average out any potential starspot crossings, we combine
the 11 2009 transits and 6 2010 transits into two lightcurves, one for each
season. We fold the lightcurves with the
\citet{2012ApJ...755...42V} period of 0.44143 days and then combine the
observations into 1-minute bins.  However, in averaging away the
potential influence of stellar activity, we introduce additional error into each
photometric measurement.  We account for that additional error by increasing
the size of our 1-$\sigma$ errors on measured parameters based on the reduced
$\chi^2$, but that correction may not fully account for stellar activity
variability.

We show the resulting photometry for the 2009
transits in Figure \ref{figure:2009_fit}, and for the 2010 transits in Figure
\ref{figure:2010_fit}.  The two lightcurves are distinctly different.  Moreover, this difference in 
shape is evident when comparing individual transits (\emph{i.e.}, not the folded
and binned versions) as well \citep[see Figure 6 from][]{2012ApJ...755...42V}. 

The 2009 transit is both shallower in depth and longer in duration than the 2010
transit.  That combination is bizarre.  Transits can conceivably change in
duration (Transit Duration Variations, or TDVs) over time due to periapsis
precession \citep{2008MNRAS.389..191P}, nodal precession
\citep{2002ApJ...564.1019M}, or perturbation from moons
\citep{2009MNRAS.392..181K} and other planets in the system
\citep{2013arXiv1304.4283N}.  In fact, nodal precession from stellar oblateness
has already been detected in one system, KOI-13, from TDVs
\citep{2012MNRAS.421L.122S}.

TDVs would also be expected to associate with changes in transit depth.  Shorter
duration transits imply a higher transit impact parameter, $b$, with a transit
chord closer to the stellar limb.  The stellar disk darkens near the limb as a
result of limb darkening.  Therefore a planet transit that evolves to shorter
duration should also have a shallower depth.

But the \planet~lightcurve gets shorter \emph{and deeper}.  That combination
implies a planet transiting nearer the stellar limb, but with that limb being
brighter than the center of the stellar disk.  Limb brightening is seen in
planets with strong gaseous absorbers \citep[like reflected light from Titan when
viewed at wavelengths where methane absorbs,][]{2002AJ....123.3473Y}.  But limb
brightening cannot happen in adiabatic stellar photospheres.

The associated phenomenon of gravity darkening, however, could help to solve the
problem.  The above discussion of TDVs assumes a stellar rotation slow enough to
have a negligible effect on the disk profile --- the star can thus be treated as
spherical.  As showed by \citet{1924MNRAS..84..665V}, a star that rotates fast
enough to become oblate also shows significant variation in brightness across
its disk.  Near the stellar equator where the local effective gravity is lower
as a result of centrifugal acceleration, the atmospheric scale height is
commensurately higher.  Consequently, the
photosphere occurs at a lower pressure level at the equator than it does at the
poles, with a correspondingly lower temperature.  

The resulting gravity-darkened stellar disk shows hotter and brighter poles
along with cooler and dimmer equatorial regions.  Interferometric imaging of the
stellar disks of nearby rapidly rotating stars has empirically confirmed the
gravity darkening concept \citep[\emph{e.g.},][]{2006Natur.440..896P,
2007Sci...317..342M,2012A&ARv..20...51V}.  Gravity darkening also affects the
lightcurves of eclipsing binary stars \citep{DIHerculis}, and has been seen in
one other transiting exoplanet lightcurve \citep[KOI-13.01,][]{KOI13,2011ApJS..197...10B}.

The \planet~lightcurve could potentially be explained therefore as a planet
transiting a gravity-darkened star.  In this scenario, the longer, shallower
2009 transits result from a low transit impact parameter as the planet traverses
across the stellar equator.  The shorter, deeper transits from 2010 would then
represent more nearly grazing transits that cross near the bright stellar pole.

In order for the gravity-darkening scenario to be plausible, \star~must be
rotating sufficiently rapidly to show significant oblateness.  Late-type stars
with convective exteriors lose angular momentum over time via stellar winds
\citep[this can be used to infer stellar ages,
\emph{i.e.},][]{2011ApJ...733L...9M}.  But a very-young star like \star~would
normally be expected to be in the midst of newborn vigorous and rapid rotation.

Present evidence suggests that \star~does rotate rapidly.  A peak in the
periodogram of the \citet{2012ApJ...755...42V} photometry with a period near
that of the planetary orbit implies synchronous rotation with a period of 10.76
hours.  Spectroscopy shows rotational broadening of the stellar lines amounting
to $v\sin i=80.\pm8~\mathrm{km/s}$, broadly consistent with synchronous
rotation given the stellar radius.

The star's youth further factors in the favor of gravity darkening because of
its large radius.  Still undergoing gravitational contraction along the Hayashi
track \citep{1961PASJ...13..450H}, the star's radius is large
\citep[$1.07~\mathrm{R_\odot}$,][]{2012ApJ...755...42V} given its low mass
\citep[$0.34~\mathrm{M_\odot}$ or $0.44~\mathrm{M_\odot}$ depending on the
model,][]{2005AJ....129..907B}.  The big radius increases the centrifugal
acceleration at the equator, which is proportional to $R_*$ for a given rotation
period.  The large radius also means that the surface gravity is lower than it
would be for an M-dwarf on the main sequence.  

That lower gravity leads to greater gravity darkening differences between the
equator and pole because the local emitted photospheric flux is proportional to
$g_\mathrm{eff}=g-a_{\mathrm{centrifugal}}$.  The equator-to-pole flux ratio is
therefore equal to
\begin{equation}
\frac{F_\mathrm{equator}}{F_\mathrm{pole}}=\frac{g_\mathrm{eq}}{g_\mathrm{p}}=\frac{g-a_{\mathrm{centrifugal}}}{g} = \frac{g-R_*\omega^2}{g}
\end{equation} 
for a stellar angular rotation rate of $\omega$.  With low gravity $g$, a large
stellar radius $R_*$, and rapid rotation $\omega$, \star~would seem an excellent
candidate for gravity darkening.

Plugging in conservative values from \citet{2012ApJ...755...42V} for these
quantities ($R_*=1.07~\mathrm{R_\odot}$, $g=105~\mathrm{m/s^2}$,
$\omega=2\pi/10.76$~hrs) yields to equator-to-pole flux ratios around $0.8$. 
The poles would then be 25\% brighter than the equator.  \star's oblateness
would be $f\sim0.1$.  Not only \emph{could} \star~be gravity darkened, if the
measurements are even close to right, it \emph{must} be gravity darkened.

\section{INDIVIDUAL FITS} \label{section:individualfits}

To evaluate whether gravity darkening on \star~could plausibly be responsible
for the unusual \planet~transit lightcurves, we first fit the 2009 December and
2010 December lightcurves separately.  To fit the \planet~transit lightcurve, we
use the program \transitfitter, as developed in \citet{2009ApJ...705..683B} and
\citet{2011ApJS..197...10B}.  It uses a Levenberg-Marquardt $\chi^2$
minimization routine to drive a numerical lightcurve model.  The
numerical model computes an explicit two-dimensional integral across the portion
of the stellar disk occulted by the planet (see Section \ref{section:model}
below for more details about the \transitfitter~algorithm).

In order to serve as a robust test for gravity darkening, we conservatively
assume a worst-case stellar mass of $M_*=0.44~\mathrm{M_\odot}$ for \star.  We
adopt a planet-synchronous stellar rotation period of 0.44841~days
\citep{2012ApJ...755...42V}.  We assume a combined quadratic limb-darkening
parameter \citep[after][]{2001ApJ...552..699B} of $c_1=u_1+u_2=0.735$, as
determined from theoretical calculations by \citet{1995A&AS..114..247C}.  To
simulate the R-band photometric observations, we model a monochromatic transit
at 0.658~$\um$.  We also assume that the planet's orbit is circular.

\begin{figure}[htbp]
\epsscale{1.}
\plotone{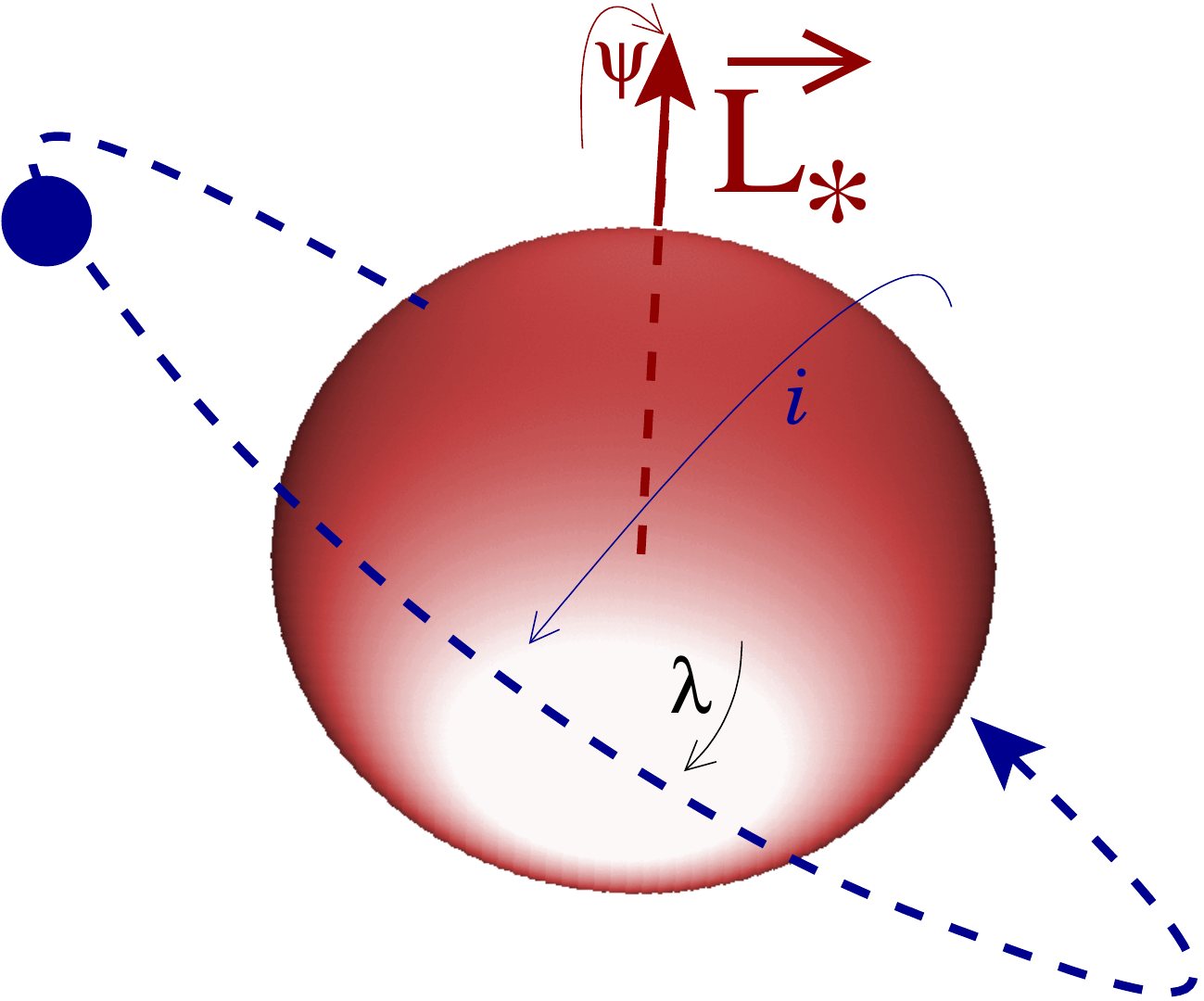}
\caption{\footnotesize Definitions of our angular geometric quantities.  The
planet's orbital inclination is $i$, measured toward the observer from the plane
of the sky.  The planet's projected spin-orbit angle
is $\asc$, as measured clockwise from stellar east.  The stellar obliquity to
the plane of the sky, $\obliq$, is measured as the angle that the north stellar
pole is tipped away from the plane of the sky. 
\label{figure:angles_geometry}}
\end{figure}

In fitting individual lightcurves, we hold limb darkening coefficient $c_1$, the
orbital period $P$, and $M_*$ constant.  We dynamically fit for $R_*$, the
planet radius $R_p$, the time at inferior conjunction $t_0$, and the
out-of-transit stellar flux $F_0$.  Spin-orbit alignment is a function of the
planetary orbit inclination $i$, the projected alignment $\lambda$, and the
stellar obliquity to the plane of the sky $\obliq$.  Figure
\ref{figure:angles_geometry} shows a diagram of these three alignment
variables.  We fit for all three of them in the case of the gravity-darkened
star, but just for $i$ in the case of no limb darkening (when changing $\lambda$
and $\obliq$ has no effect on the lightcurve).

We fit each lightcurve with both a gravity-darkened rotating star model (blue in
figures) and a conventional non-rotating spherical star model with no gravity
darkening (red in figures).  Table \ref{table:bestfit} contains the resulting
best-fit system parameters, based on angle definitions as shown in Figure
\ref{figure:angles_geometry}.  Figures \ref{figure:2009_fit} and
\ref{figure:2010_fit} show the observations, best-fit spherical and
gravity-darkened lightcurves (both from \transitfitter),  and fit residuals.

\begin{table*}[htb]
\centering
\begin{tabular}{lccccccc}
 & $\chi^2_r$ &$R_*$ ($\mathrm{R_\odot}$)&$R_p$ ($\mathrm{R_{Jup}}$)& $t_0$ (s) & $i$ & $\asc$ & $\obliq$ \\
\hline
2009 only, no gravity darkening & 2.43 & $1.00\pm0.06$ & $1.60\pm0.14$ & $30861720\pm50$  & $74^\circ\pm5^\circ$ & --- & --- \\
2009 only, with gravity darkening & 2.11 & $1.19\pm0.07$ & $2.00\pm0.17$ & $30861700\pm200$ & $64^\circ\pm3^\circ$ & $90.^\circ\pm22^\circ$ & $2^\circ\pm19^\circ$ \\
2010 only, no gravity darkening & 2.24 & $0.98\pm0.14$ & $5.4 \pm2.1$  & $60848600\pm50$  & $45^\circ\pm7^\circ$ & --- & --- \\
2010 only, with gravity darkening & 1.54 & $1.39\pm0.11$ & $1.80\pm0.20$ & $60848300\pm290$ & $58^\circ\pm5^\circ$ & $136^\circ\pm33^\circ$ & $31^\circ\pm25^\circ$ \\
2009 \& 2010, no gravity darkening & 3.03 & $1.15\pm0.04$ & $2.11\pm0.14$ & $60848560\pm70$ & $61^\circ\pm2^\circ$ & --- & --- \\
\end{tabular}
\caption{\footnotesize Best-fit values for fits of the 2009 and 2010
\planet~transit lightcurves, with and without gravity darkening.  Angles $i$, $\asc$, and $\obliq$ are as shown
in Figure \ref{figure:angles_geometry}.  The epoch time $t_0$ is measured in 
seconds past 2009 January 1 at midnight UTC.}
\label{table:bestfit}
\end{table*}

\subsection{2009 Transit} 

In 2009 (Figure \ref{figure:2009_fit}), the transit bottom is flat --- or
possibly even convex.  A conventional spherical star fit (\emph{i.e.}, one
without gravity darkening, shown in red) cannot reproduce this convexity at
all.  The gravity-darkened star fit (in blue) does a reasonable job of
reflecting the transit bottom convexity by having the planet transit chord
perpendicular to the stellar equator.  We measure that angle as the projected
spin-orbit alignment angle $\asc$ where $\asc$ is measured clockwise from
stellar east.  The star itself has a low obliquity to the plane of the sky,
$\obliq$ (where $\obliq=i-90^\circ$ with $i$ being the conventional stellar
inclination relative to the line of sight).  

\begin{figure}[htbp]
\epsscale{1.}
\plotone{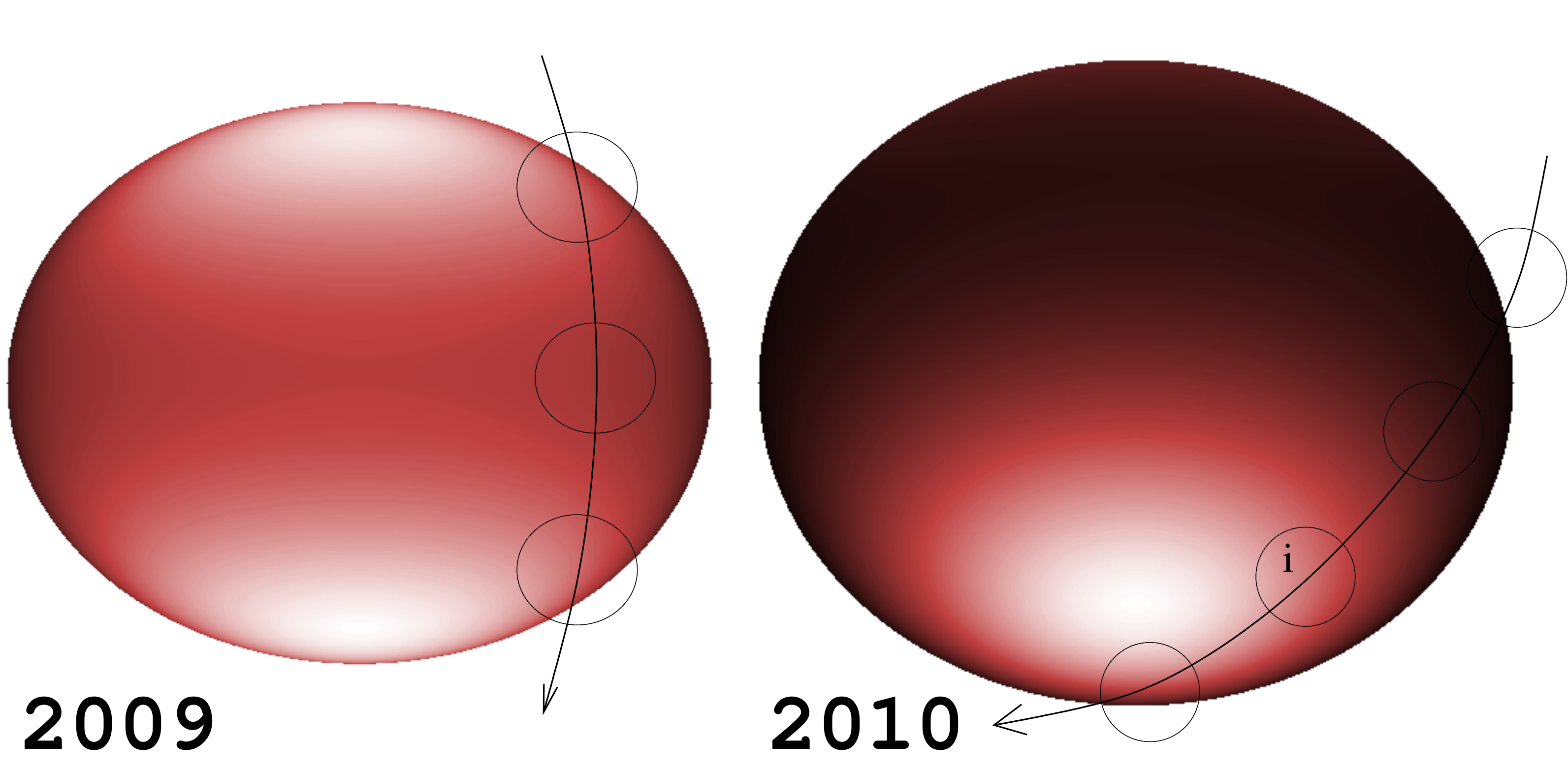}
\caption{\footnotesize Transit geometry of the best-fit gravity-darkened models
for 2009 (left) and 2010 (right) photometry of \star.  The two images are to
scale, accounting for different best-fit stellar radii in the two cases (see
Table \ref{table:bestfit}).  The larger best-fit radius in 2010 leads to more
severe gravity darkening at the stellar equator.  In each image the planet's
projected path is shown as a series of appropriately scaled black circles
separated in time by 2400 seconds (40 minutes).  The apparent curvature of the
projected paths is real: it derives from tracking of the full three-dimensional
planetary orbit trajectory in this unusual case where the planet's orbit is less
than 2 stellar radii in semimajor axis.  The center planet circle from 2009
corresponds to inferior conjunction.  In 2010 the circle with the tiny mark in
it (an `i') denotes inferior conjunction --- an oblique transit path across a 
gravity-darkened, oblate star leads to the long transit duration and asymmetric
lightcurve evident in the photometric data (Figure \ref{figure:2010_fit}).  
\label{figure:both_indiv_geometry}}
\end{figure}

We show a graphical representation of the transit geometry in the
left subfigure
in Figure \ref{figure:both_indiv_geometry}.  By transiting perpendicular to the
oblate stellar equator, the total transit duration is shorter than the
equivalent $\asc=0$ transit by a factor of $1-f$ where $f$ is the stellar
oblateness.

The planet first transits the oblate star at relatively high stellar latitude. 
Nearer the pole the stellar photosphere is hotter, and therefore the stellar
flux is higher.  This effect is compensated by the countervailing effect
of limb darkening, however, leading to total planet-occulted fluxes that are not
too different from those at mid-transit when the planet covers the cooler (and
dimmer) equator.  

If anything, the gravity-darkened model underfits the intensity of the `horns' on
the lightcurve that lead to the transit bottom convexity.  We could exaggerate
the horns in the fit by increasing the gravity darkening parameter $\beta$
\citep{1924MNRAS..84..665V}, which we fix at $\beta=0.25$.  However based on the
overall accuracy of the data, the error for which is significantly increased by
stellar spots and flares, we elect not to modify $\beta$ at this time to avoid
overfitting.  Additional observations can reduce the overall noise level and may
permit a measurement of $\beta$ in the future.

The lightcurve is nearly symmetric for this $\lambda=90^\circ$ transit, with the
small best-fit asymmetry provided by the slight stellar obliquity in this fit
(north pole pointed $2^\circ$ away from the observer).  The relatively high
errors for the spin-orbit angles $\asc$ and $\obliq$ result from the model's
ability to fit the flat transit bottom near-equally well by decreasing
(increasing) the projected angle $\lambda$ and increasing (decreasing) the
stellar obliquity $\obliq$.  That degeneracy also leads to greater uncertainty
in the time of inferior conjunction $T_0$.  With an oblate star, transits with
projected alignments $\lambda$ that differ from $0^\circ$, $90^\circ$,
$180^\circ$, and $270^\circ$ have mid-transit times that do not coincide with
the time of inferior conjunction, as might usually be assumed.

Note the arced projected planet trajectory in Figure
\ref{figure:both_indiv_geometry}.  It is real.  Because the planet's semimajor
axis is so tiny --- less than 2 stellar radii -- approximating the planet's
path as straight no longer provides sufficient fidelity.

\subsection{2010 Transit}

The 2010 lightcurve (Figure \ref{figure:2010_fit}) has a significantly different
shape to that from 2009.  Instead of a flat transit bottom, the transit bottom
is sharply peaked.  It shows no convexity.  And it is decidedly asymmetric, with
a long tail toward the ingress side.

The spherical-star (non-gravity-darkened) model does a particularly poor job of
fitting the 2010 transit.  Specifically, our model is unable to reproduce the
long ingress tail.  In fitting for the high transit bottom curvature (the `V'
shape) without gravity darkening, the model runs off the rails toward a grazing
transit for an unreasonably large planet radius ($R_\mathrm{p} =
5.4\pm2.1~\mathrm{R_{Jup}}$).

On the other hand, the gravity-darkened fit reproduces the 2010 lightcurve well.  We show the best-fit gravity-darkening transit
geometry at right in Figure \ref{figure:both_indiv_geometry}.  

The precise shape of the ingress tail drives the fit toward a higher stellar
radius than the 2009 fit:  $1.39~\mathrm{R_\odot}$ (2010) as opposed to 
$1.19~\mathrm{R_\odot}$ (2009).  The $1.39~\mathrm{R_\odot}$ matches precisely
the spectroscopically-derived value from \citet{2005AJ....129..907B}.  However,
given that the analysis of the spectroscopy did not account for the
gravity-darkened nonuniformity of emission across the stellar disk, the
similarity in stellar radius values is likely coincidence.  Slight changes to
the tail shape, as might be present but swamped by stellar noise, could allow
for smaller stellar radii.

The larger inferred stellar radius ($1.39~\mathrm{R_\odot}$ vs.
$1.19~\mathrm{R_\odot}$) observed in 2010 than in 2009 drives more
severe gravity darkening at the stellar equator.  The fit's sharp transit bottom
derives from a transit path that crosses near the hot but small stellar polar
region.  The long total transit duration results from a transit that starts at
the stellar equator and maximizes its interior path using the inherent
orbital curvature.  The long tail then arises due to an early ingress along the
cold and dim stellar equator.

The radius discrepancy may owe to the inherent noisiness of the
lightcurve as driven by stellar activity.  In particular the 2010 tail at
ingress could be an artifact resulting from an imperfect fit to the
out-of-transit stellar variation.  In order to resolve the radius discrepancy,
we need to simultaneously fit both lightcurves so as to force the fit into
coherence.  Such a simultaneous fit requires an understanding of how the transit
geometry could evolve from that seen in 2009 to that seen in 2010.  Precession
of the stellar spin and planetary orbit angular momenta could provide such a
mechanism.

\section{PRECESSION} \label{section:precession}

The gravity-darkened best-fit values for the 2009 and 2010 transits show
reasonable agreement, given that they were fit separately.  The stellar and
planetary radii overlap within $2\sigma$.  The spin-orbit parameters $\asc$ and
$\obliq$, however, are very different.  Taking into account the highly oblate
star and tiny planetary orbit semimajor axis (just $1.7~\mathrm{R_*}$), though,
the two measurements could be reconciled if the planet-star system
experienced precession in the intervening year.

\subsection{Form of Precession}\label{section:precessionform}

\begin{figure}[!tbhp]
\epsscale{1.}
\plotone{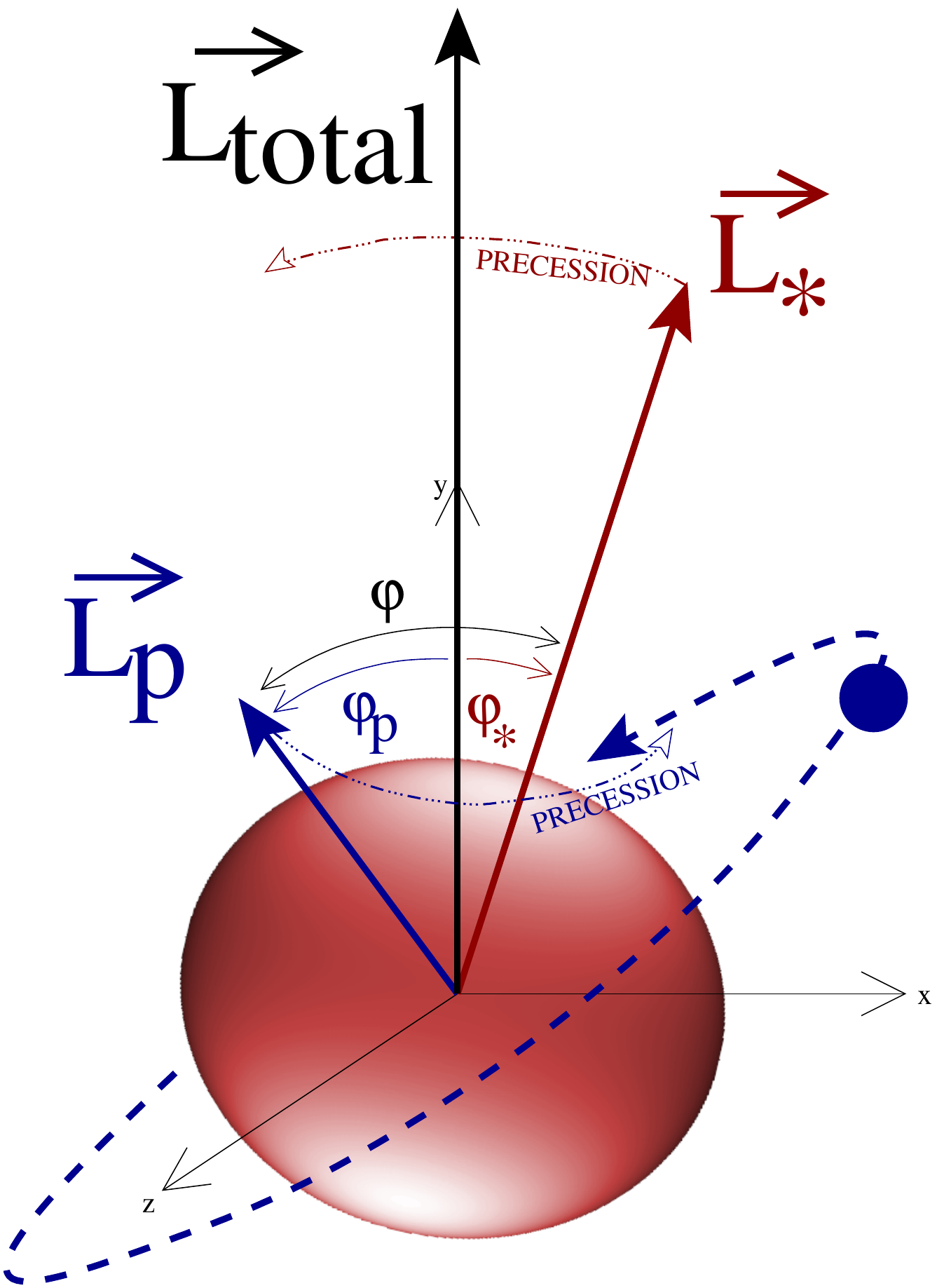}
\caption{\footnotesize Precession geometry for the case of an oblate star (red)
orbited by a single planet (blue).  Both the stellar spin angular momentum
$\Lspinvec$ and the planetary orbit angular momentum $\Lorbitvec$ precess around
the system's invariable plane, defined by $\Ltotalvec$.  In the case where
$\Lspin \gg \Lorbit$ the geometry reduces to $\Lorbitvec$ precessing around
$\Lspinvec$, as for the International Space Station precessing around Earth, for
example.  When $\Lorbit \gg \Lspin$, then $\Lspinvec$ effectively precesses 
around $\Lorbitvec$, as would happen in a Sun-Jupiter system in the absence of
other bodies.  In the case of \planet, $\Lorbitvec \sim \Lspinvec$, and the more
complex geometry described in this figure is required, where 
$\Lorbitvec$ and $\Lspinvec$ both precess around the net
angular momentum of the system $\Ltotalvec$. The precession arrows show the direction of `positive' precession in
the mathematics.  In reality the precession is negative, i.e. retrograde, or
clockwise as seen from above the stellar north pole.  
\label{figure:precession_geometry}}
\end{figure}

In the case of a two-body system consisting of an oblate star and a close-in
planet, torques between the planet and the star's rotational bulge induce nodal
precession (Figure \ref{figure:precession_geometry}).  Similar precession
scenarios are familiar within the solar system, where the Earth's spin and orbit
both precess.  

Earth's rotation axis precesses around the plane containing the
Sun every 26,000 years, resulting from torques applied to its rotational bulge
by the Sun and the Moon.  This process maintains Earth's obliquity (axis tilt),
changing only the azimuthal direction in which the axis points\footnote{At least
it would in a simple Earth-Sun or Earth-Moon-Sun system.  The real Earth's
obliquity actually \emph{can} vary, due to an interaction between the precessions of
Earth's spin and orbit resulting from torques from the other planets (see
\citet{1993Natur.361..615L,moonless.Earth}).}.

Earth's orbit is inclined by $1.57^\circ$ with respect to the solar system's
invariable plane --- the plane normal to the solar system's net angular momentum
vector $\Ltotalvec$.  Earth's orbital angular momentum vector $\Lorbitvec$
precesses around $\Ltotalvec$ every $\sim100,000$ years.  Jupiter's gravity
provides the dominant torque driving the precession of Earth's orbit, though,
not the Sun's rotational bulge.

The case of \planet~is more complex than that of the Earth's precessions owing
to the similarity of the magnitudes of the stellar spin angular momentum 
$|\Lspinvec|\equiv\Lspin = \mathbb{C} M_* R_*^2 \omega$ and the planetary orbit angular
momentum $|\Lorbitvec|\equiv\Lorbit = M_p a_p^2 n$, where $\mathbb{C}$ is the moment of inertial
coefficient (0.4 for a uniform-density sphere, 0.059 for the Sun), $M_*$ is the
stellar mass, $R_*$ is the stellar radius, $\omega$ is the stellar rotation rate
(in, say, radians per second), $M_p$ is the planetary mass, $a_p$ is the
planetary orbital semimajor axis, and $n$ is the planet's orbital mean motion
(again in radians per second).  We assume a circular orbit here for simplicity. 
\citet{2012MNRAS.421L.122S} investigated a similar case where 
$\Lorbit \sim \Lspin$ in the case of exoplanet KOI-13.01.
The ratio of the
system angular momentum represented by the planet's orbit to that in the stellar
spin is
\begin{equation}
\frac{\Lorbit}{\Lspin}=\frac{1}{\mathbb{C}}\frac{M_p}{M_*}\frac{n}{\omega}
\bigg(\frac{a}{R_*}\bigg)^2~~~~.
\end{equation}
Using parameters from \citet{2005AJ....129..907B} and \citet{2012ApJ...755...42V}, as shown in our Table
\ref{table:PTF1}, the possible values of  $\frac{\Lorbit}{\Lspin}$ for the
\planet~system range from 0.080 if $M_p=1.0~\mathrm{M_{Jup}}$ to 0.45 if 
$M_p=5.5~\mathrm{M_{Jup}}$ (the \citet{2012ApJ...755...42V} radial velocity
upper limit).

\begin{table}[tbh]
\centering
\begin{tabular}{|r|l|}
\hline \hline
$M_*$ & $0.34~\mathrm{M_\odot}$ or  $0.44~\mathrm{M_\odot}$\\ 
$R_*$ & $1.39~\mathrm{R_\odot}$ \\ 
$\mathbb{C}$ & 0.059  \\ 
$\omega=n$ & $0.44841\pm0.00004$~days \\ 
$a_p$ &  $1.80~\mathrm{R_\odot}$  \\ 
$M_p$ & $\leq 5.5\pm1.4~\mathrm{M_{Jup}}$ \\ 
\hline \hline
\end{tabular}
\caption{\footnotesize PTF-1 system parameters, from \citet{2005AJ....129..907B}
and \citet{2012ApJ...755...42V}.The value for the stellar moment of inertia
coefficient $\mathbb{C}$ is assumed to be that of the Sun.
\label{table:PTF1}}
\end{table}

Hence, in the case of \planet, the stellar spin pole and the planetary orbit
normal both precess around their mutual net angular momentum vector as shown in
Figure \ref{figure:precession_geometry}.  To quantify the geometry, we define
$\spinorbit$ to be the angle between $\Lspinvec$ and $\Lorbitvec$.  This
value, the spin-orbit angle, is constant as a function of time (assuming that
there are no other objects in the system that affect it on the same timescale). 
We then separately define individual obliquities $\spinorbit_{p}$ and $\spinorbit_*$ to be
the angular distances between $\Lorbitvec$ and $\Ltotalvec$, and between
$\Lspinvec$ and $\Ltotalvec$ respectively.  The two sum to $\spinorbit$,
\begin{equation}\label{eq:spinorbitsum}
\spinorbit_p+\spinorbit_*=\spinorbit~~,
\end{equation}
with 
\begin{equation}\label{eq:spinorbitsines}
\frac{\sin{\spinorbit_p}}{\sin{\spinorbit_*}}=\frac{\Lspin}{\Lorbit}~~~.
\end{equation}

The mutual precession of $\Lorbitvec$ and $\Lspinvec$ is driven by the torque
$\vec{\tau}$ between the planet and the stellar rotational bulge.  The rates of
change of the angular momentum vectors $\frac{d\Lorbitvec}{dt}$ and
$\frac{d\Lspinvec}{dt}$ are equal and opposite according to Newton's third
law: 
\begin{equation} 
\vec{\tau} \equiv \frac{d\Lorbitvec}{dt}=-\frac{d\Lspinvec}{dt}~~~. 
\end{equation}
The magnitude of the torque in the case where $\Lorbit \sim \Lspin$ is 
identical to that in the endpoint cases when $\Lorbit \gg \Lspin$ or 
$\Lorbit \ll \Lspin$.  But in the $\Lorbit \sim \Lspin$ case, instead of 
$\Lorbitvec$ precessing around
in a circle with a total circumference of $2\pi \Lorbit \sin{\spinorbit}$ 
(as it does when 
$\Lorbit \ll \Lspin$), it must instead traverse a distance of 
$2\pi \Lorbit \sin{\spinorbit_p}$.  Therefore given the precession rate  
$\dot{\Omega}_p$ for 
the longitude of the ascending node of the planet's orbit 
in the simpler  $\Lorbit \ll \Lspin$ case,
then the full mutual precession rate $\dot{\Omega}$ for the $\Lorbit \sim \Lspin$ is
\begin{equation}\label{eq:precessionrate}
\dot{\Omega} = \dot{\Omega}_p \frac{\sin{\spinorbit}}{\sin{\spinorbit_p}} ~~~.
\end{equation}
The precession rate for the stellar rotation pole 
must be the same, \emph{i.e.}
\begin{equation}
\frac{\dot{\Omega}_*}{\sin{\spinorbit_*}}=\frac{\dot{\Omega}_p}{\sin{\spinorbit_p}}
\end{equation}
since they precess together, and
\begin{equation}\label{eq:precessionrate2}
\dot{\Omega} = \dot{\Omega}_* \frac{\sin{\spinorbit}}{\sin{\spinorbit_*}}~~.
\end{equation}

Equation \ref{eq:precessionrate} has some interesting consequences in the
$\Lorbit \sim \Lspin$ case.  First of all, for $\spinorbit<90^\circ$, the full
precession rate is always faster than the orbit precession rate from the
$\Lorbit \ll \Lspin$ case, $\dot{\Omega}>\dot{\Omega}_p$.  In the limit that
$\spinorbit$ is small, the full precession rate can be approximated as
$\dot{\Omega}=\dot{\Omega}_p\frac{\Ltotal}{\Lspin}$ --- hence a factor of two
increase in the equal-angular-momentum case $\Lorbit = \Lspin$.  However, in a
situation where the planet orbits retrograde,  $90^\circ<\spinorbit<180^\circ$,
it is possible to have $\sin{\spinorbit_p} > \sin{\spinorbit}$, in which case
slower, and in some cases extremely slow, precession is possible.  For example,
take the case where $\Lorbit = \Lspin$, and $\spinorbit=176^\circ$.  Here the
torque is low, but $\spinorbit_p$ is $88^\circ$ and $\Lorbitvec$ must precess
all the way around.  The result of this thought experiment would be a precession
rate 14 times slower than in the $\Lorbit \ll \Lspin$ case, all else being
equal.

For the expected values for $\frac{\Lorbit}{\Lspin}$, full precession rates for
\planet~should be between 10\% and 50\% faster than they would be if we had
assumed $\Lorbit \ll \Lspin$.  Therefore, detection of precession could constrain
the planet's mass both by the precession rate and by the relative amplitude of
the stellar and orbital precessions.

\subsection{Rate of Precession}

The precession rate in systems with $\Lorbit \sim \Lspin$ can be calculated
using either Equation \ref{eq:precessionrate} or Equation
\ref{eq:precessionrate2}.  Both equations, however, derive from the separate
precession rates that are valid for systems with asymmetric angular momentum. 
In theory, those rates ($\dot{\Omega}_p$ and $\dot{\Omega}_*$) can be calculated
exactly given the stellar mass $M_*$, the planetary mass $M_p$, the planet's
orbital mean motion $n\equiv2\pi/P$ (where $P$ is the orbital period), and
knowledge of the stellar interior structure.

In a more conventional system where $\Lspin \gg \Lorbit$, as for the precession
of one of Saturn's moons for instance, an approximation of 
the precession rate can be written
as \citep{SolarSystemDynamics}  
\begin{equation}\label{equation:planetprecessionfull}
\dot{\Omega}_p=-n\cos{\spinorbit}
\bigg[\frac{3}{2}J_2\bigg(\frac{R_*}{a}\bigg)^2-
\frac{27}{8}J_2^2\bigg(\frac{R_*}{a}\bigg)^4-
\frac{15}{4}J_4\bigg(\frac{R_*}{a}\bigg)^4
\bigg]~~~
\end{equation}
to order 4 in $R_*/a$.  Here $J_2$ is
the stellar rotation-driven quadrupole moment and $n$ is the planet's orbital mean
motion.
This is typically approximated as
\begin{equation}\label{equation:planetprecession}
\dot{\Omega}_p=-n\cos{\spinorbit}\frac{3}{2}J_2\bigg(\frac{R_*}{a}\bigg)^2
\end{equation}
when $R_p/a$ is small.
Similarly, the stellar precession rate in the opposite limiting case where 
$\Lorbit \gg \Lspin$ might be given as
\begin{equation}\label{equation:starprecession}
\dot{\Omega}_* = -\frac{3}{2}J_2\frac{GM_p\cos{\spinorbit}}{a^3n\mathbb{C}}~~~~.
\end{equation}

\begin{figure*}[tb]
\epsscale{1}
\plotone{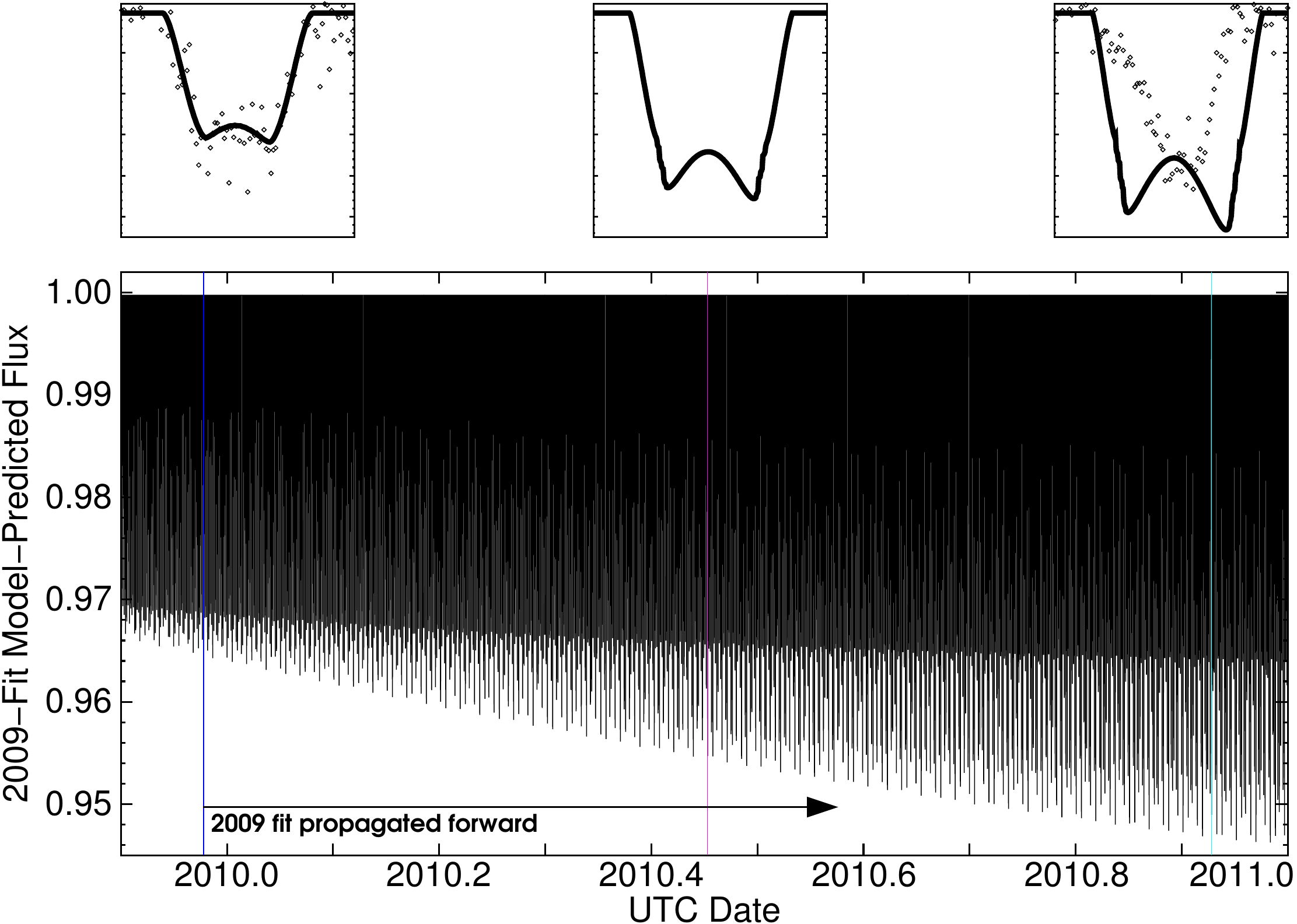}
\caption{\footnotesize  Forward-extrapolation of the individual fit to the 2009
December photometry.  The bottom plot shows the predicted photometric flux seen
from \star~over the course of the year between the 2009 and 2010 observations. 
This Figure uses as its initial conditions those from the 2009-only fit from
Table \ref{table:bestfit}.  We show zoom-ins of what individual transits look
like inset at top.  The colored vertical lines show the times that correspond to
the inset transits at top.  The best-fit projected alignment value of
$\asc=90^\circ$ leads to unusually slow precession for the precise values from
Table \ref{table:bestfit}.  However other valid sets of transit parameters with
$\asc$ further from $90^\circ$ but still within the error bars would see 
substantially faster precession rates.  Note that data points above 
flux values
of 1.002 have been clipped in the insets so that the ranges correspond with the
plot at bottom.
\label{figure:2009_individualfit_precessed}}
\end{figure*}

A complicating factor is that for \planet, $R_*/a$ is not small at up to 0.77 for
$R_*=1.39~R_\odot$
(since $a=1.8~\mathrm{R_\odot}$).  Therefore the
higher order terms ($(R_*/a)^4$ and higher) may substantially contribute.  Using
the point-core assumption (quite  good for stars), the stellar $J_2$ is
\begin{equation}\label{equation:J2}
J_2=\mathbb{C}f
\end{equation}
where $f$ is the stellar oblateness, defined as
\begin{equation}
f\equiv\frac{r_\mathrm{equatorial}-r_\mathrm{pole}}{r_{\mathrm{equatorial}}}~~.
\end{equation}
Using the \star~parameters as reported in \citet{2012ApJ...755...42V} and shown
in Table \ref{table:PTF1}, we calculate
that $f=0.20$.  Then, incorporating an assumption that the stellar moment of
inertia coefficient is $\mathbb{C}=0.059$ (similar to that of the Sun), 
we calculate that $J_2=0.012$ using Equation \ref{equation:J2}.  For spin-orbit
angles $\spinorbit$~near alignment ($\spinorbit$ small), the first term
in Equation \ref{equation:planetprecessionfull} is
$\dot{\Omega}_p=1.6\times10^{-6}$~radians/s, corresponding to a precession
period of just 45 days!  This value is consistent with a different calculation by 
\citet{2012ApJ...755...42V} that yielded a period of ``tens to hundreds of 
days".

The second-leading term in Equation \ref{equation:planetprecessionfull} yields
just $2.6\times10^{-8}$~radians/s.  Therefore despite $R_*/a$ not being small,
the low $J_2$ driven by central mass concentration in the star
($\mathbb{C}=0.059$) allows us to treat the precession rate as in Equation
\ref{equation:planetprecession} to within a few percent.  We therefore adopt
Equation \ref{equation:planetprecession} for the remainder of the present study.

An additional complication can arise from distortions to the stellar
gravitational field due to the tidal bulge induced by the planet.  A more
careful calculation of $\dot{\Omega}_p$ could take into account the higher order
$J_n$ terms with $n\ge4$, with $J_n$ calculated numerically as suggested by
\citet{2012ApJ...756L..15H}.  The size of this bulge should vary for an inclined
planet as the stellar radius changes with sub-planetary latitude.  This effect
in combination with the non-prolate nature of the tidal bulge due to the
planet's proximity \citep[as for HAT-P-7, ][]{2012ApJ...751..112J} might
necessitate a complete numerical calculation of the average potential around the
planet's orbit in order to most accurately determine $\dot{\Omega}_p$.

\begin{figure*}[bt]
\epsscale{1}
\plotone{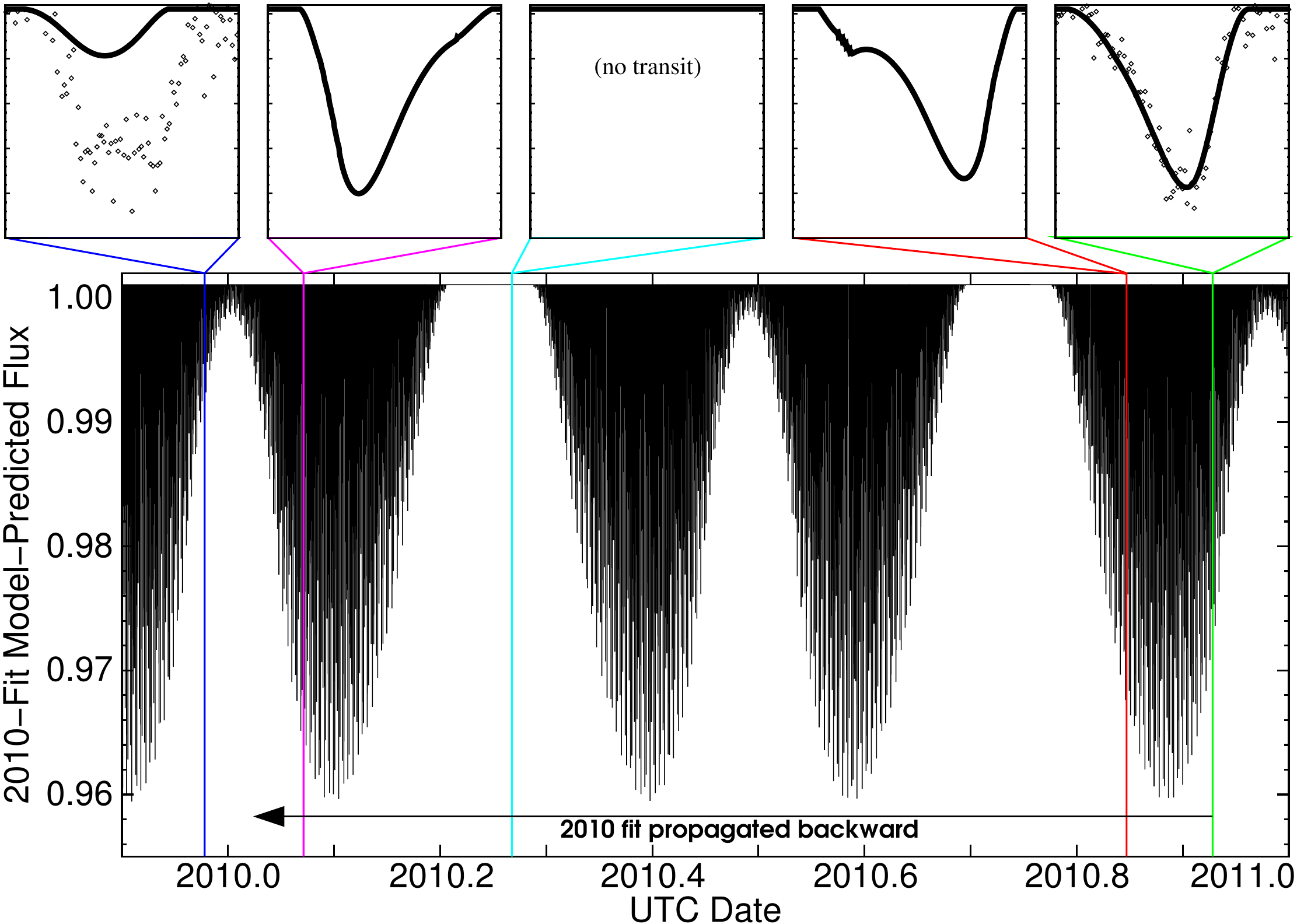}
\caption{\footnotesize Backward-extrapolation of the individual fit for the 2010
December photometry.  As in Figure \ref{figure:2009_individualfit_precessed},
the bottom shows a plot of the transit photometry in the time between the 2009
and 2010 observations.  The insets at top depict what individual transit events
look like at the times shown with colored vertical bars.  With a precession period of 179 days, this extrapolation
comes close to providing a reasonable explanation for \planet's
transit lightcurve changes.  Particularly interesting is the possibility that
the transits may disappear for some portion of the precession cycle.
\label{figure:2010_individualfit_precessed}}
\end{figure*}

\section{MODEL} \label{section:model}

The core of the \transitfitter~lightcurve algorithm is a two-dimensional
numerical integration in polar coordinates across the occulted portion of the
stellar disk to obtain observed stellar flux.  Since that explicit integration
is relatively slow (a single fit takes about a day or so to complete), for
coarse fits we also use the mostly analytical approximation from
\citet{2002ApJ...580L.171M}, Section 5.  That approximation does not contain a
separate formula for the case when $z\le p$ (\emph{i.e.} when the planet covers
the point at the center of the stellar disk --- we use here the \citet{2002ApJ...580L.171M}
variable definitions that $p\equiv R_p/R_*$ and $z\equiv d/R_*$ where $d$ is the
projected separation between the center of the planet and the center of the star).  For non-rotating, spherical stars,
the usual formula for the stellar flux $I^*(z)$ (valid for $z<1-p$) works fine, since the stellar
disk is nearly uniform at the geometric center.  But for fast-rotating stars, a
separate case is needed (parallel to Case 9 in the \citet{2002ApJ...580L.171M}
analytical case):
\begin{equation*}
I^*(z) = \frac{p^2+2pz-z^2}{2p^2(p+z)^2}\int_0^{z+p}I(r)2r\mathrm{d}r
\end{equation*}
\begin{equation}
 +\frac{(p-z)^2}{2p^2(p-z)^2}\int_{z-p}^0I(r)2r\mathrm{d}r~~,
\end{equation}
with the radial direction of the integral (the direction matters 
for gravity-darkened stars since
the disk is non-isotropic) being in the direction from the star center toward 
the planet center.  Negative values in the second term indicate an integral in
the opposite direction.

The second change that we have made to the algorithm is the incorporation of
precession of both the planetary orbit and the stellar spin, as described
theoretically in Section \ref{section:precession}.  The new routine has two
parts:

\paragraph*{Preprecession:}  Whenever there is a change in the projected
planetary orbit alignment $\asc$, the orbital inclination $i$, the stellar
obliquity with respect to the plane of the sky $\obliq$, the stellar mass $M_*$,
the planetary mass $M_p$, the orbital period $P$, the planetary orbital
eccentricity $e$, or the stellar rotation period $P_\mathrm{rot_*}$,
\transitfitter~calculates time-independent quantities in a process that we call
`preprecession'.  In preprecession, \transitfitter~executes the following steps:
\begin{enumerate}\setcounter{enumi}{-5}
\item Calculate $\LorbitvecN$, the planet orbit angular momentum vector in 
the sky frame at the time of epoch.
\item Calculate $\LspinvecN$, the stellar spin angular momentum in the sky frame
at the time of epoch.
\item Calculate $\LtotalvecN=\LorbitvecN+\LspinvecN$.
\item Transform $\LorbitvecN$ and $\LspinvecN$ into a coordinate system with
$\LtotalvecN$ along the $z$-axis as in Figure \ref{figure:precession_geometry}
using Euler angles.
\item Calculate $\dot{\Omega}$ from Equation \ref{eq:precessionrate}.
\end{enumerate}

\paragraph*{Precession:}  Then, if either the time $t$ or the epoch $t_0$ 
change, \transitfitter~calculates the precession using the results of the 
preprecession calculations:
\begin{enumerate}
\item Calculate $\Lorbitvec(t)$ and $\Lspinvec(t)$ in the $\Ltotalvec$ frame 
with a rotation of both through an angle $\dot{\Omega}t$.
\item Transform $\Lorbitvec(t)$ and $\Lspinvec(t)$ back into the sky frame using
Euler angles.
\item Update the values for $\asc$, $i$, $\obliq$, and the observationally
irrelevant azimuthal angle of the projected stellar spin pole with respect to
sky north.
\end{enumerate}

\section{EXTRAPOLATION} \label{section:extrapolation}

Using the technical background for precession from Section
\ref{section:precession} implemented in \transitfitter~as described in
Section \ref{section:model}, we can now look at whether precession may be
responsible for the different \star~lightcurves seen in 2009 and 2010.  To
start, we take the best-fit individual solutions from Section
\ref{section:individualfits} and extrapolate forward (2009) and backward
(2010) as a test case for the plausibility of the precession hypothesis.  For
now we assume that $M_\mathrm{p}=1.0~\mathrm{M_{Jup}}$ with regard to precession
rate and $\spinorbit_\mathrm{p}$ calculations.

Figure \ref{figure:2009_individualfit_precessed} shows the extrapolation of the
2009 fit.  It is clearly not close to being able to explain the 2010 data.  The
projected alignment $\asc$ near $90^\circ$ is the reason for its failure.  Near
$\asc=90^\circ$ (and $270^\circ$), the planet is in a polar orbit around the
star.  Hence there is no torque to drive precession forward.  The result is a
super-long precession period of 8486 days (23 years).  Any real value for
$\lambda$ would have to be significantly different from $90^\circ$ to be
consistent with precession.  However, because of the high
degeneracy between $\asc$ and $\obliq$ for the 2009 fit and the resulting
uncertainties in these fitted parameters, such a situation remains plausible.

Extrapolating from the 2010 individual fit yields a much more interesting and
plausible result (Figure \ref{figure:2010_individualfit_precessed}).  For the
2010 individual best-fit parameters the precession period is 179 days (in the
retrograde direction, and assuming $M_\mathrm{p}=1.0~\mathrm{M_{Jup}}$).  Over
the course of those 179 days the transit depth increases, then decreases but
does not quite drop to zero before rebounding to the same maximum depth.  After
the second depth peak, the depth decreases until the transits disappear for about
a month.  Then the cycle repeats.

The shape of the transits also varies during the precession.  During the first
half of the cycle the planet transits across the cool stellar equator first,
before then crossing at or near the pole.  This was the situation for the 2010
December lightcurve.  These transits are asymmetric with the deepest part closer
to egress.  At mid-cycle the transits are shallow and near-symmetric.  In the
later half of the cycle, the transits are mirror images of those in the first
half, transiting the pole first and then the darker low stellar latitudes. 
Hence in the second half of the cycle the transits are asymmetric with the
deepest part on the ingress side.

Note also the shift of mid-transit time for the extrapolated 2009 December data
relative to the timing of the actual observed transit.  The shift results from
the shift in the time of the deepest part of the transit relative to the timing
of inferior conjunction (see Figure \ref{figure:both_indiv_geometry}).

Taking the 2009 individual best-fit and extrapolating forward cannot replicate
the 2010 transit because of a slow rate of precession that allows for only small
changes in $i$, $\obliq$, and $\lambda$ in the intervening year.  The 2010
individual best-fit cannot reproduce the 2009 data when propagated backwards
either, owing to the wrong stellar obliquity $\obliq$, planetary inclination
$i$, and projected alignment $\lambda$ when evolving the system following the
algorithm described in Section \ref{section:model}.  The precession-propagated
values for the alignment variables $\obliq$, $i$, and $\lambda$ are rather
sensitive to both the initial values and the planet mass $M_p$.  Therefore it is
possible that using slightly different 2010 initial conditions --- equally
likely given the uncertainties for those values in Table \ref{table:bestfit} ---
along with a different value for $M_p$, we could obtain a coherent,
self-consistent precession to simultaneously model both lightcurves.

\section{JOINT FIT} \label{section:jointfit}

\begin{figure*}[tbhp]
\epsscale{1.15}
\plotone{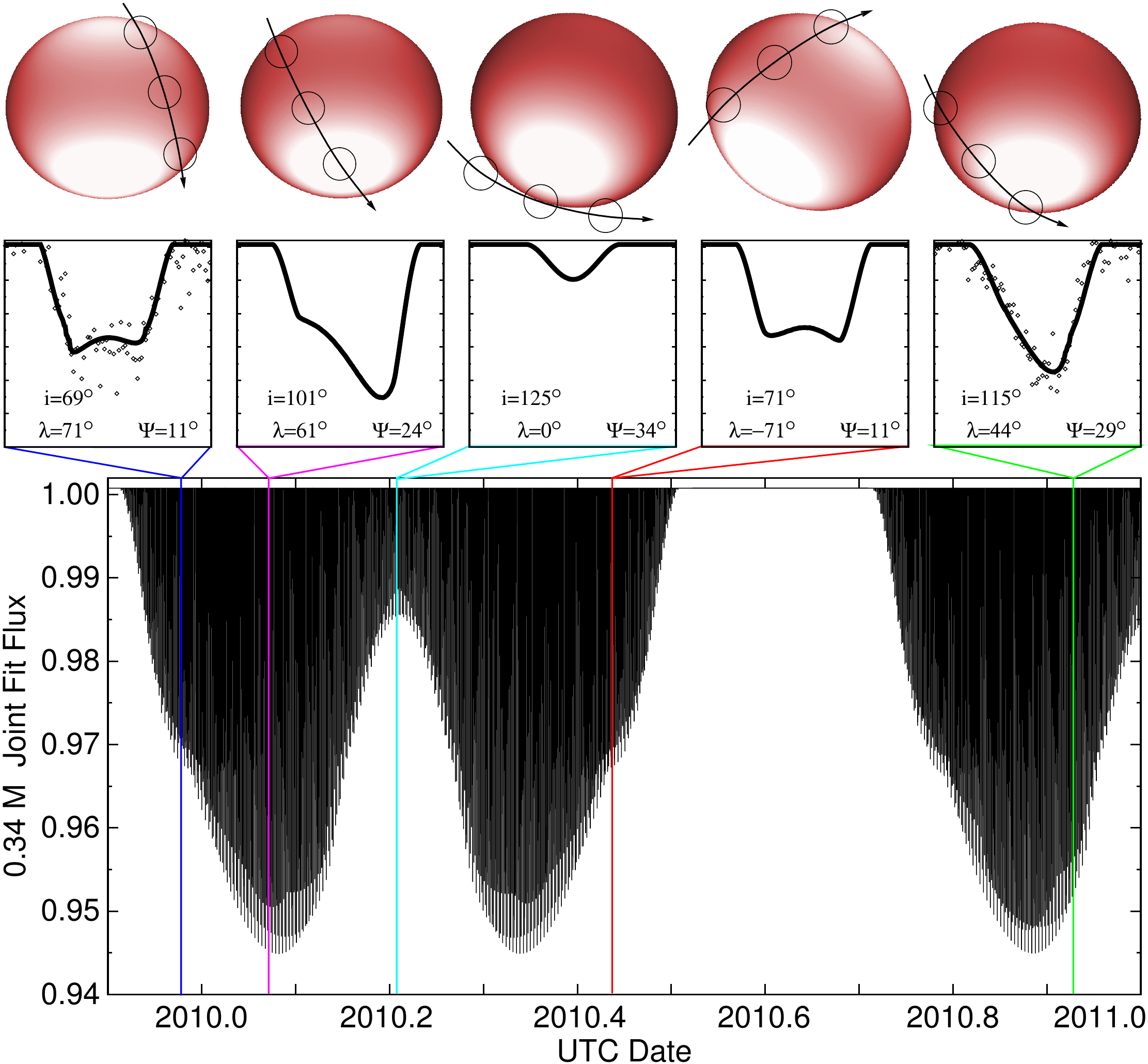}
\caption{\footnotesize This figure shows the best-fit self-consistent joint fit
to both the 2009 and 2010 photometric data under the assumption that the stellar
mass $M_*=0.34~\mathrm{M_\odot}$.  The insets at top show the model lightcurves,
observed points (open diamonds), and synthetic images with the planet's
trajectory at 5 different epochs between 2009 and 2010.  The time of the 2009
observational photometry from \citet{2012ApJ...755...42V} is at far left, and the 
2010 photometry is the inset at far right.  The middle three insets show transit
lightcurve shapes and graphic depictions of what the transit might have looked
like at three different times between the 2009 and 2010 observations, as
predicted by the fit parameters from Table \ref{table:jointfits}.  The bottom
graph shows the model output over 1.1 years from 2009.9 through 2011.0 UTC, with
the times of the insets at top denoted with colored vertical lines.  As in the
case of the 2010 individual fit shown in Figure 
\ref{figure:2010_individualfit_precessed}, this joint fit predicts
periods during which the planet does not transit at all during the course of the
system's precession.
\label{figure:0_34}}
\end{figure*}

\begin{figure*}[tbhp]
\epsscale{1.15}
\plotone{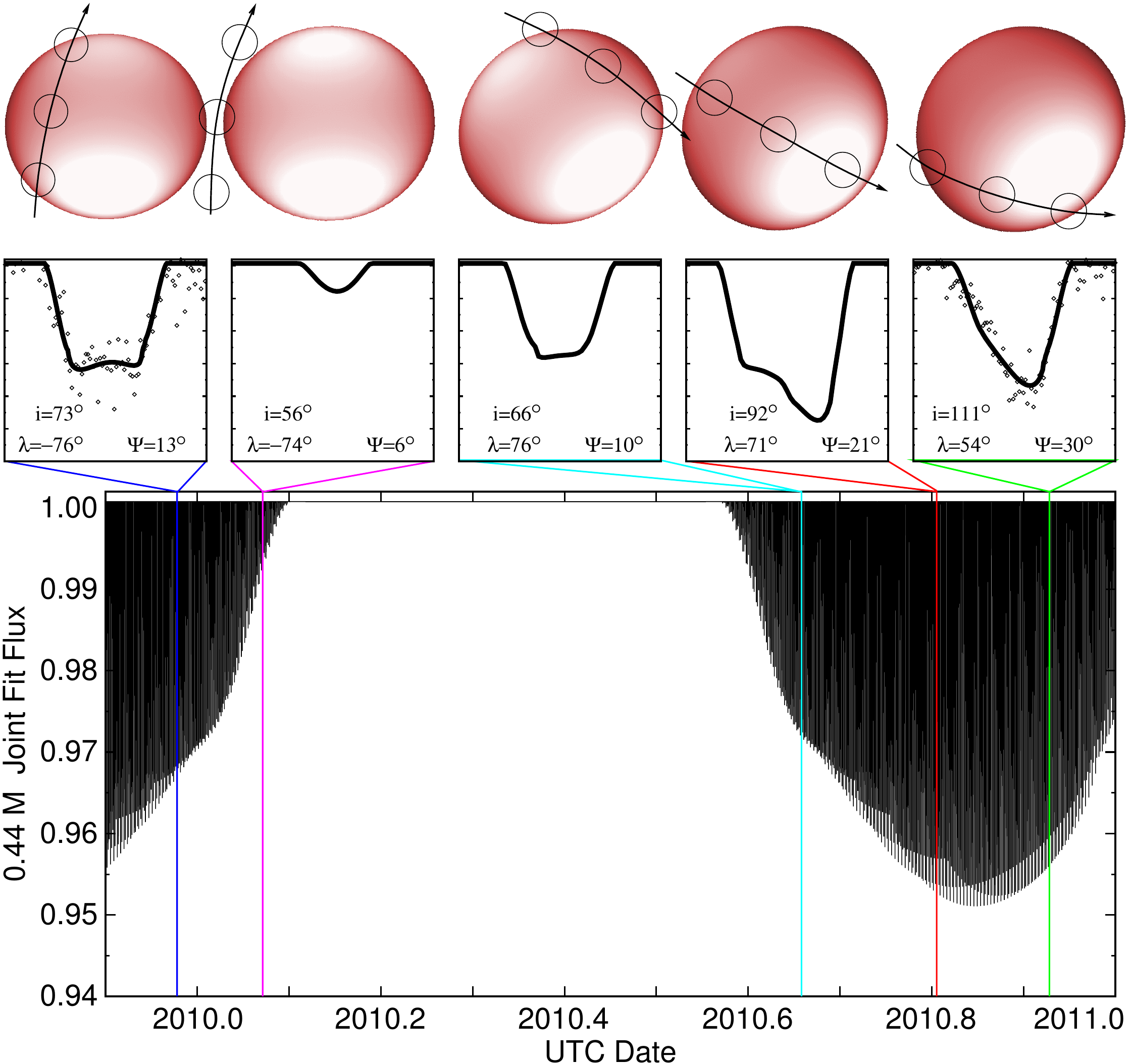}
\caption{\footnotesize This figure shows the best-fit self-consistent joint fit
to both the 2009 and 2010 photometric data under the assumption that the stellar
mass $M_*=0.44~\mathrm{M_\odot}$.  The insets at top show the model lightcurves,
observed points (open diamonds), and synthetic images with the planet's
trajectory at 5 different epochs between 2009 and 2010.  The time of the 2009
observational photometry from \citet{2012ApJ...755...42V} is at far left, and the 
2010 photometry is the inset at far right.  The middle three insets show transit
lightcurve shapes and graphic depictions of what the transit might have looked
like at three different times between the 2009 and 2010 observations, as
predicted by the fit parameters from Table \ref{table:jointfits}.  The bottom
graph shows the model output over 1.1 years from 2009.9 through 2011.0 UTC, with
the times of the insets at top denoted with colored vertical lines.  This
$M_*=0.44~\mathrm{M_\odot}$ solution precesses more slowly than the
$M_*=0.34~\mathrm{M_\odot}$ solution in Figure \ref{figure:0_34}, resulting in
less than a full precession cycle between the 2009 and 2010 observational
lightcurves.
\label{figure:0_44}}
\end{figure*}

A complete joint fit of both the 2009 and 2010 observations, accounting for
precession and fitting for the planetary orbital period, may be able to
comprehensively explain the both the lightcurve shapes and their variability.
After an extensive trial-and-error search we were able to identify
self-consistent sets of conditions that yield satisfactory simultaneous joint
fits to both the 2009 and 2010 lightcurves.  The ranges explored included
$0.8~R_\odot<R_*<1.6~R_\odot$, $0<M_p<100~\mathrm{M_{Jup}}$, and initial (2010)
values for $i$, $\obliq$, and $\lambda$ within the bounds of the 2010 individual
fit and equivalents in both prograde and retrograde directions.  

With just two epochs, however, the fits are not necessarily unique.  We describe
two of them here, one for the assumption of $M_*=0.34~\mathrm{M_\odot}$ and the
other for $M_*=0.44~\mathrm{M_\odot}$, each of the spectroscopically-derived
stellar masses described in \citet{2005AJ....129..907B}.

\begin{table}[htb]
\centering
\begin{tabular}{r|c|c|}
\multicolumn{3}{c}{\textbf{Parameters for Joint Fits}}\\
\hline
\hline 
                 &$0.34~\mathrm{M_\odot}$       & $0.44~\mathrm{M_\odot}$  \\ \hline
$R_*$            &$1.04\pm0.01\mathrm{R_\odot}$ & $1.03\pm0.01\mathrm{R_\odot}$ \\
$R_p$            &$1.64\pm0.07\mathrm{R_{Jup}}$ & $1.68\pm0.07\mathrm{R_{Jup}}$ \\
$P$              &$0.448410\pm0.000004$ days    & $0.448413\pm0.000001$ days \\
$t_0$            &$60848500\pm100~\mathrm{s}$   & $60848363\pm38~\mathrm{s}$ \\
$i$              &$114.8^\circ\pm1.6^\circ$     & $110.7^\circ\pm1.3^\circ$ \\
$\lambda$        &$43.9^\circ\pm5.2^\circ$      & $54.5^\circ\pm0.5^\circ$ \\
$\obliq$         &$29.4^\circ\pm0.3^\circ$      & $30.3^\circ\pm1.3^\circ$ \\
$M_p$            &$3.0\pm0.2~\mathrm{M_{Jup}}$  & $3.6\pm0.3~\mathrm{M_{Jup}}$ \\
$\spinorbit$     &$69^\circ\pm3^\circ$          & $73.1^\circ\pm0.6^\circ$ \\
$\spinorbit_*$   &$18^\circ$                    & $20.2^\circ$ \\
$\spinorbit_p$   &$51^\circ$                    & $52.9^\circ$ \\
$P_{\dot{\Omega}}$&$-292.6$~days                & $-581.2$~days \\
$f$              &0.109 & 0.083\\
$\chi_r^2$       &2.17 & 2.19 \\
\hline \hline
\end{tabular}
\caption{\footnotesize Best-fit parameters from the self-consistent, joint fit
of the 2009 and 2010 \citet{2012ApJ...755...42V} lightcurves.  Epochs $t_0$ are
measured in seconds after 2009 January 1 00:00 UTC (JD 2454832.5).  The orbital 
period is $P$.
\label{table:jointfits}}
\end{table}

\begin{table}[htb]
\centering
\begin{tabular}{r|r|r|}
\multicolumn{3}{c}{\textbf{Back-Propagated Alignment Parameters}}\\
\hline
\hline 
                 &$0.34~\mathrm{M_\odot}$       & $0.44~\mathrm{M_\odot}$  \\ \hline 
$t_0$				  & 30861500 s       & 30861370 s \\
$i$              &$69.1^\circ$      & $72.7^\circ$ \\
$\lambda$        &$71.1^\circ$      & $-76.1^\circ$ \\
$\obliq$         &$10.7^\circ$      & $12.8^\circ$ \\
\hline \hline
\end{tabular}
\caption{\footnotesize Alignment parameters from the self-consistent, joint fit
of the 2009 and 2010 \citet{2012ApJ...755...42V} lightcurves as propagated
back to the time of the 2009 transit.  Our model generates the same lightcurve
using these as its initial values as it does using the values at the 2010 epoch
shown in Figure \ref{table:jointfits}
\label{table:jointfitpropagations}}
\end{table}

Table \ref{table:jointfits} shows the best-fit parameters for each case: 
$0.34~\mathrm{M_\odot}$ and $0.44~\mathrm{M_\odot}$.  In each case, in addition
to the fit parameters, which use the mid-transit time of the 2010 observations
as their epoch, we show in Table \ref{table:jointfitpropagations} the precessed 
values as propagated back to the time
of the 2009 observations.  A graphical representation of the observing geometry
as precessed along with transit lightcurves and their evolution in each case are
shown in Figures \ref{figure:0_34} and \ref{figure:0_44}.

We use the covariance matrix from the Levenberg-Marquardt fitting algorithm to
generate the quoted errors in Table \ref{table:jointfits}, modified by a 
correction for the high $\chi^2$ of the final fit.  The $\chi^2$ higher than 1.0
results from accurate photometry of an inherently variable star due to surface
activity (starspots).  We treat the variability as a source of red noise over
and above the photometric precision, compensating for it by multiplying the
formal covariance errors by $\sqrt{\chi^2}$.  This compensation approach is an 
approximation due to the non-Gaussianity of the stellar variability.  A more
sophisticated approach like residual permutation \citep[\emph{\'a
la}][]{2009ApJ...693..794W} is not warranted given the degeneracy between our
two different solutions.  

Interestingly, the uncertainties on fit parameters coming out of the
joint fit are significantly tighter than the uncertainties from fitting each
transit individually.  For example, the measured uncertainty on the projected
alignment $\asc$~was $33^\circ$ when fitting the 2010 transit individually, and
$25^\circ$ for the stellar obliquity $\obliq$.  But when fitting for the 2010
transit along with the 2009 transit and including precession, those uncertanties
plummet to $5.2^\circ$ and $0.3^\circ$ respectively!  What's going on here?  

It turns out that the requirement that the 2010 initial conditions propagate
backward into the 2009 conditions via precession constrains the system more
tightly than do the transit geometries necessary to generate the lightcurve
shapes by themselves.  With the complex systemic precession as described in
Section \ref{section:precession}, the initial conditions in 2010 must propagate
into the conditions that replicate the 2009 transit.  This requirement very
tightly constrains the initial values for $\asc$ and $\obliq$, for instance.  It
also affects the planet mass $M_p$ via the partition of the full spin-orbit
alignment angle $\spinorbit$ into $\spinorbit_\mathrm{p}$ and $\spinorbit_*$. 
If the planet's mass is too small, then it is unable to pull the star around
into the orientation required for the other transit.  If the planet's mass is
too big, then it can pull the star around too much.  Similarly, in order for the
system to arrive in the proper orientation \emph{at the right time}, the
precession period directly constrains the combination of $R_*$, $M_p$, and
$\spinorbit$.

Essentially these constraints somewhat resemble those for asteroids on a
collision course with Earth.  Even with uncertain knowledge of an asteroid's
present-day orbital parameters, if you were to know that it was going to collide
with the Earth at a certain time in the future, that would by itself give you
much more powerful knowledge of what its present-day parameters must be even
without better present-day observations.  And similar to the asteroid analogy,
the further separated in time the target is from the present, the better those
constraints will be.  Thus future observations of \planet~transits should be
capable of driving parameters to such precision that the ultimate uncertainties
will be dominated by systematic errors instead of measurement error.

Different precession periods $P_{\dot{\Omega}}$ characterize the two independent
solutions.

\subsection{Stellar Mass $M_*=0.34~\mathrm{M_\odot}$ Case}  In the $M_*=0.34~\mathrm{M_\odot}$
case, the precession period is 292.6 days (in the retrograde direction, hence
the negative sign in Table \ref{table:jointfits}).  The system therefore
undergoes 1.25 precessions in between the 2009 and 2010
\citet{2012ApJ...755...42V} lightcurves.  The spin-orbit misalignment angle of
$\spinorbit=69^\circ$ makes this a highly inclined planet orbit.  But with
$\spinorbit=69^\circ$ instead of the $\spinorbit=90^\circ$ (as was the case for
the 2009-only fit from Section \ref{section:individualfits}), the precession rate
$\dot{\Omega}$ is much faster.

We found this solution by trying various acceptable fits to the 2010 lightcurve
alone, then looking at the back-propagation to 2009 and trying to get close
enough for the Levenberg-Marquardt solver to zero in on a fit.  As such the
$M_*=0.34~\mathrm{M_\odot}$ joint fit reproduces the 2010 lightcurve with
similar geometry to the individual fit from Section
\ref{section:individualfits}.  Without precession transit lightcurves around
gravity-darkened stars leave a four-fold geometric degeneracy \citep[see][Figure
3]{2011ApJS..197...10B}.  While the specific geometry shown in Figure
\ref{figure:both_indiv_geometry} indicates a retrograde orbit geometry, the
joint fit uses the prograde equivalent of the same geometry.

In numerous searches we were not able to find satisfactory fits under any
retrograde solution.  This does not necessarily mean that such fits do not
exist, only that we did not find one.  However, given the effort that we
employed in searching, retrograde solutions may very well be ruled out.

In this joint solution, the late 2010 planet initially transits a cool and dim
near-equatorial region at ingress, and then the hot bright pole near egress to
reproduce the lightcurve asymmetry in the 2010 lightcurve.  In contrast to the
individual fit, however, the joint fit does a relatively poorer job of fitting
the long tail present in the data at ingress.  A smaller stellar radius in the
joint fit explains the difference.  

All of our attempts at a joint fit with $R_*\sim1.4\mathrm{R_\odot}$, which
would have matched both the 2010 individual fit and the spectroscopic estimate,
failed.  The short duration of the 2009 transit and its sharp ingress and egress
prevent a satisfactory solution for larger stars.  This is only true under our
assumption of a circular orbit for the planet, of course --- if the planet's
orbit were eccentric
\citep[\emph{e.g.}][]{2007PASP..119..986B,2008ApJ...679.1566B,2008ApJ...678.1407F},
then a solution that matches the spectroscopic radius might still be possible. 
We did not pursue such a solution, but the addition of more photometric epochs
from future observations might allow constraints on orbital eccentricity.

The $R_*=1.04\mathrm{R_\odot}$ stellar radius from the
$M_*=0.34~\mathrm{M_\odot}$ joint fit has other consequences, as well.  The
smaller radius means a smaller $v\cos{\obliq}$ (``$v\sin{i}$"), using our
assumed synchronous stellar rotation period.  It also leads to less gravity
darkening and a lower stellar oblateness.

Gravity darkening breaks the azimuthal degeneracy on the stellar disk, which
allows us to determine which part of the star the planet crosses in transit. 
The broken symmetry leads to our measured planetary inclination of $114.8^\circ$
--- unusual, considering that inclinations have heretofore always been
$0^\circ\leq i \leq 90^\circ$.  By convention, this inclination greater than
$90^\circ$ indicates that the planet is over the star's southern hemisphere at
inferior conjunction.

When precessed backward to the epoch of the 2009 observations, this fit
replicates the flatness of the transit bottom by first transiting near the hot
bright north pole, but while that pole is tilted away from the observer.  As it
travels across the stellar disk to the south, it also moves further from the
location of the projected stellar axis, leaving it relatively further from the
south pole on egress.  By being farther from the brighter pole on egress and
closer to the less bright pole on ingress, the transit bottom overall is fairly
flat.

The planet's mass $M_p$ drives the overall precession rate $\dot{\Omega}$ and
thus the relative placement of the 2009 observation within the precession
sequence.  All else being equal, higher planetary masses drive faster
precessions for prograde orbits by reducing $\spinorbit_*$, as per Equation
\ref{eq:precessionrate2}.  In \transitfitter, these small changes in $M_p$
squeeze or extend the precession plot at the bottom of Figure \ref{figure:0_34}
like an accordion.  

The planet's mass also affects the partition of the spin-orbit alignment angle
$\spinorbit$ into the the planet precession angle $\spinorbit_p$ and the stellar
precession angle $\spinorbit_*$ as described by Equations \ref{eq:spinorbitsum}
and \ref{eq:spinorbitsines}.  Hence, for larger changes in the planet's mass,
again all else being equal, the shape and character of the transits during
precession change as $\spinorbit_p$ and $\spinorbit_*$ change.

Our best-fit value for planet mass in the $M_*=0.34~\mathrm{M_\odot}$ case is
$M_p=3.0~\mathrm{M_{Jup}}$.  For masses significantly different than
$M_p=3.0~\mathrm{M_{Jup}}$, on the order of $0.0~\mathrm{M_{Jup}}$ or
$8.0~\mathrm{M_{Jup}}$, the transit shapes change such that flat bottoms do not
occur at any time during the precession.  For smaller changes in $M_p$,
different values of $M_p$ push the flat-bottomed transits to times incompatible
with the 2009 observations.

Note, however, that flat-bottomed transits also occur at another point in the
precession cycle.  At the time of the red line in Figure \ref{figure:0_34} (the
inset second from right), the transit signatures also have flat bottoms in such
a way that could match the shape of the 2009 data.  Assuming a zero-mass (test
particle) planet extends the precession to bring this second flat-bottom
location closer to the 2009 observation epoch, but not all the way there. 
Faster precession could bring another instance of this second flat-bottom
location forward from a new cycle.  However, doing so requires a planet mass so
large that the precession character alters, and the flat-bottomed portion no
longer exists.

Although the observations cannot be explained using the second flat-bottomed
area under the assumption that $M_*=0.34~\mathrm{M_\odot}$, our second solution
using $M_*=0.44~\mathrm{M_\odot}$ does use its equivalent.

\begin{figure}[tbhp]
\epsscale{1.}
\plotone{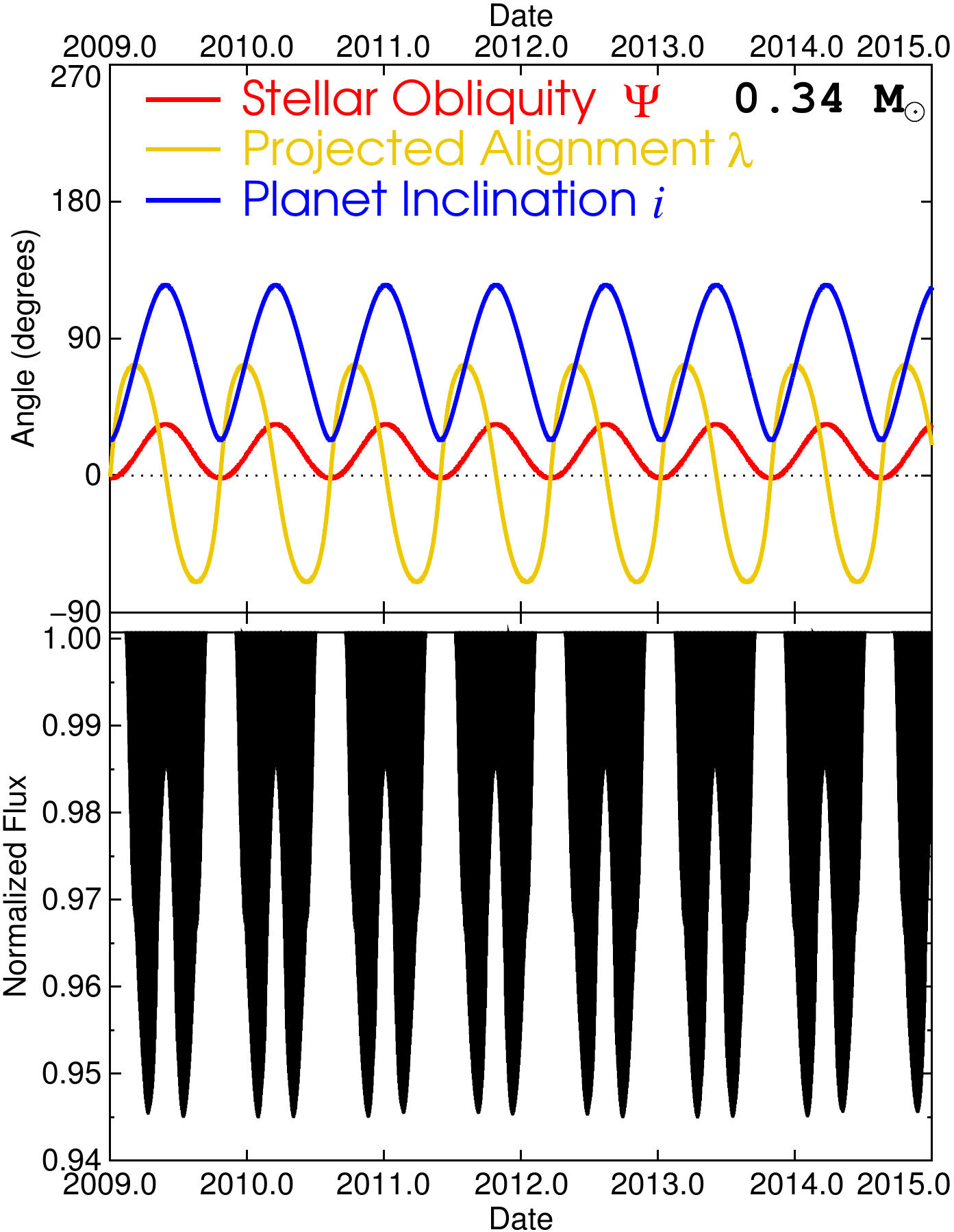}
\caption{\footnotesize In this figure we plot the future predicted state of the
\planet~system from the $M_*=0.34~\mathrm{M_\odot}$ model.  We use black to
indicate future transit depth, red to indicate future stellar obliquity to the
plane of the sky $\obliq$, yellow for future projected spin-orbit alignment
$\lambda$, and blue to show future planetary orbital inclination $i$.
\label{figure:0.34_forwardprojection}}
\end{figure}

\subsection{Stellar Mass $M_*=0.44~\mathrm{M_\odot}$ Case}  As in the $M_*=0.34~\mathrm{M_\odot}$
case, with $M_*=0.44~\mathrm{M_\odot}$ we were unable to fit both observations
using a stellar radius near the spectroscopic value of $1.4~\mathrm{R_\odot}$. 
Instead, our best fit in the $M_*=0.44~\mathrm{M_\odot}$ was with a smaller
star:  $R_*=1.03~\mathrm{R_\odot}$, similar to the best-fit value in the
$M_*=0.34~\mathrm{M_\odot}$ case.  Thus, the precession period for this
$M_*=0.44~\mathrm{M_\odot}$ is longer, owing to a similar radius but higher
stellar mass.  This leads to higher gravity which results in a less oblate
stellar figure (lower $J_2$) and therefore lower precession torques and a slower
precession rate (see Equations \ref{equation:planetprecession} or
\ref{equation:starprecession}).

The overall geometry for the 2010 epoch is similar to that in the
$M_*=0.34~\mathrm{M_\odot}$ fit above (see Figure \ref{figure:0_44}, rightmost
inset).  The projected alignments $\lambda$ are similar, as are the stellar
obliquities.  As a result, the total spin orbit alignment $\spinorbit$ is
similar as well.  This is essentially the same solution as that for
$M_*=0.34~\mathrm{M_\odot}$, except that it uses a conjugate version of the 
flat-bottomed portion of the precession.  Due to the slower precession, this
solution reproduces the 2009 data with the opposite projected alignment
$\lambda$, resulting in a very similar model lightcurve result.

With slower precession, the $M_*=0.44~\mathrm{M_\odot}$ fit shows a more
extended period without transits, in this case fully 6 months long.

The reduced chi squared $\chi_r^2$ for each fit is similar:  2.17 for
$M_*=0.34~\mathrm{M_\odot}$ and 2.19 for $M_*=0.44~\mathrm{M_\odot}$.  Both are
significantly above 1.0.  As a pre-main-sequence M-dwarf, \star~is particularly
noisy, which presumably drives the $\chi_r^2$ of our fits to be higher than the
photon shot noise ideal.

\section{FUTURE PROJECTION}\label{section:projection} 

\begin{figure}[tbhp]
\epsscale{1.}
\plotone{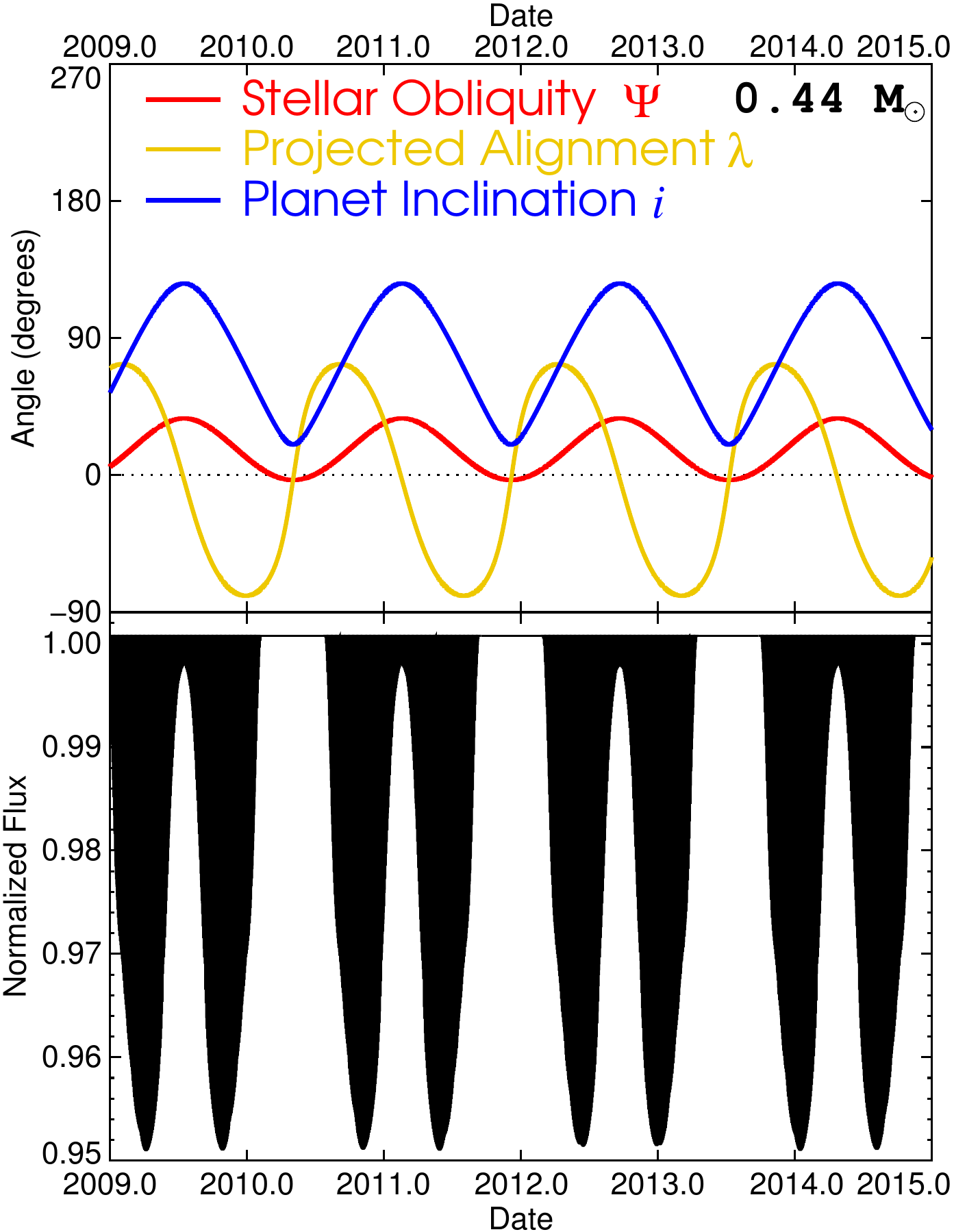}
\caption{\footnotesize Same as Figure \ref{figure:0.34_forwardprojection}, 
but for $M_*=0.44~\mathrm{M_\odot}$.
\label{figure:0.44_forwardprojection}}
\end{figure}

The joint fit offers several avenues for testing the veracity of the precessing
gravity-darkened model.  Figures \ref{figure:0.34_forwardprojection} and
\ref{figure:0.44_forwardprojection} show projections until the end of 2014 for
transit depth, planet orbit inclination $i$, the projected spin-orbit alignment
$\lambda$, and the stellar obliquity to the plane of the sky $\obliq$.

\paragraph*{Disappearing Transits}  Both joint fit models predict periods of
several months during which no transits should occur.  At these times, the
planet's impact parameter $b=a\cos(i)/R_*$ is greater than $1+\frac{R_p}{R_*}$,
as can be inferred from the planetary orbital inclination, which we plot in blue
in Figures \ref{figure:0.34_forwardprojection} and
\ref{figure:0.44_forwardprojection}.  Future photometric campaigns to observe
the \planet~transit could confirm our model if they were to show that the
transits disappear.  Although the times will shift with varying $M_*$, the
no-transit periods last for 0.21 years with $M_*=0.34~\mathrm{M_\odot}$ and for
0.46 years with $M_*=0.44~\mathrm{M_\odot}$.  They recur each cycle, starting at
$2010.51+NP_{\dot{\Omega}}$ for $M_*=0.34~\mathrm{M_\odot}$ and
$2010.10+NP_{\dot{\Omega}}$ for $M_*=0.44~\mathrm{M_\odot}$ where $N$ is any
integer.  Non-detections of transits would also place tight constraints on the
initial conditions, helping, for instance, to nail down $M_*$ or to rule out
one of our two $M_*$ scenarios.

\paragraph*{Changing Orbit Inclination}  Related to the disappearance of
transits, the predictions for \planet's changing orbital inclination could be
tested directly by radial velocity (RV) measurements.  Sets of RV measurements
spanning the planet's entire orbital phase would be required, each separated in
time by several months.  The planet's 10.8-hour orbital period makes it possible
to potentially span an orbit in a single night of observing.  However, \star's
large apparent magnitude and high inherent stellar noise would both be
problematic for the RV approach.

\paragraph*{Changing Stellar Inclination}  Our models predict that the star's
rotation pole should be precessing around the net angular momentum vector of the
system, as shown in Figure \ref{figure:precession_geometry}.  Both fits show
that this stellar pole precession should result in changes to the stellar
obliquity as measured from the plane of the sky ($\obliq$).  We show our
predictions for changing stellar obliquity as the red curves in Figures
\ref{figure:0.34_forwardprojection} and \ref{figure:0.44_forwardprojection}. 
The obliquity should vary between $\sim0^\circ$ and $\sim30^\circ$. 
Spectroscopy of \star~over several months' time could show this changing stellar
obliquity using variations in stellar line rotational widths.  The line widths
should vary as $v\cos(\obliq)$  --- ``$v\sin(i)$" under the conventional
definition for $i$ as stellar obliquity to the line of sight.  Due to the cosine
dependence, however, the variation in $v\cos(\obliq)$ should only be $\sim13\%$.

\paragraph*{Changing Stellar Spectrum}  As the stellar inclination changes, the
star presents to Earth more or less of its hot polar regions.  When viewed more
nearly pole-on, the star should have a spectrum with a higher effective
temperature and an earlier spectral type than when more of the equator is
visible.  This effect should also lead to long-term changes in the overall
magnitude of the star.  It should appear brighter when the pole is presented to
Earth, and dimmer when the pole is more nearly in the plane of the sky.

\paragraph*{Changing Projected Alignment}  Nodal precession causes large changes
in the projected spin-orbit alignment of the system, $\lambda$, which we show in
yellow in Figures \ref{figure:0.34_forwardprojection} and
\ref{figure:0.44_forwardprojection}.  The alignment should vary between
$\sim-80^\circ$ and $\sim+70^\circ$ and should be zero in between during the
period of grazing transits (\emph{i.e.} the second inset from left in Figure
\ref{figure:0_44}).  These changes in $\lambda$ could be measured from the
Rossiter-McLaughlin effect
\citep[\emph{e.g.}][]{1924ApJ....60...15R,1924ApJ....60...22M,2012ApJ...757...18A}
or from using stroboscopic starspots
\citep[\emph{e.g.}][]{2011ApJS..197...14D,2011ApJ...740L..10N,2011ApJ...743...61S}.
Practitioners of either method should take care to account for the apparently
curved path of the planet across the stellar disk that arises from the very
small orbital semimajor axis, as shown in the synthetic images at the top of
Figures \ref{figure:0_34} and \ref{figure:0_44}.

\paragraph*{Changing Transit Shapes}  A combination of the previous effects
leads to changes in the specific transit geometry as a function of time.  Those
changes manifest as variations in the shape of the planet's transit lightcurve,
seen in the inset lightcurves at the top of Figures \ref{figure:0_34} and
\ref{figure:0_44}.  Continued photometric monitoring of the type already done
by \citet{2012ApJ...755...42V} can confirm the nodal precession of \planet,
differentiate between the two models that we have presented, and allow for
precise measurements of all transit parameters, including $M_*$.

\begin{figure}[tbhp]
\epsscale{1.}
\plotone{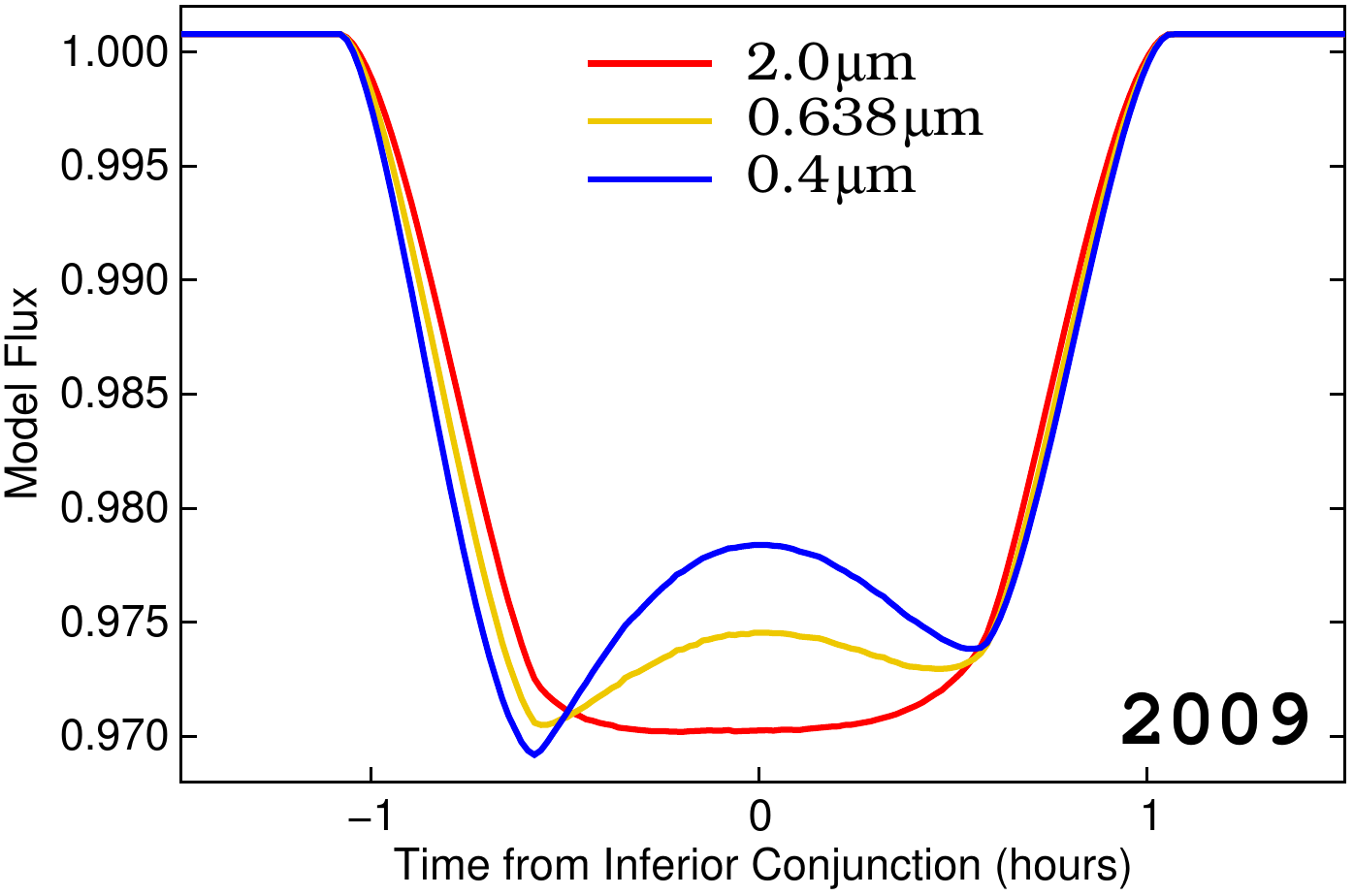}
\caption{\footnotesize This figure shows fluxes for the 
$M_*=0.34~\mathrm{M_\odot}$ model at three different wavelengths:  $2.0~\um$
(red), $0.638~\um$ (yellow), and $0.4~\um$ (blue).  These three models all assume
an identical limb-darkening parameter $c_1=0.735$.  Usually transits are nearly
achromatic, with different fluxes in different wavebands resulting only from
different limb darkening.  With gravity darkening, however, the transit
lightcurve should look substantially different at different wavelengths owing to varying
temperatures across the gravity-darkened stellar disk.
\label{figure:2009_colors}}
\end{figure}

\begin{figure}[tbhp]
\epsscale{1.}
\plotone{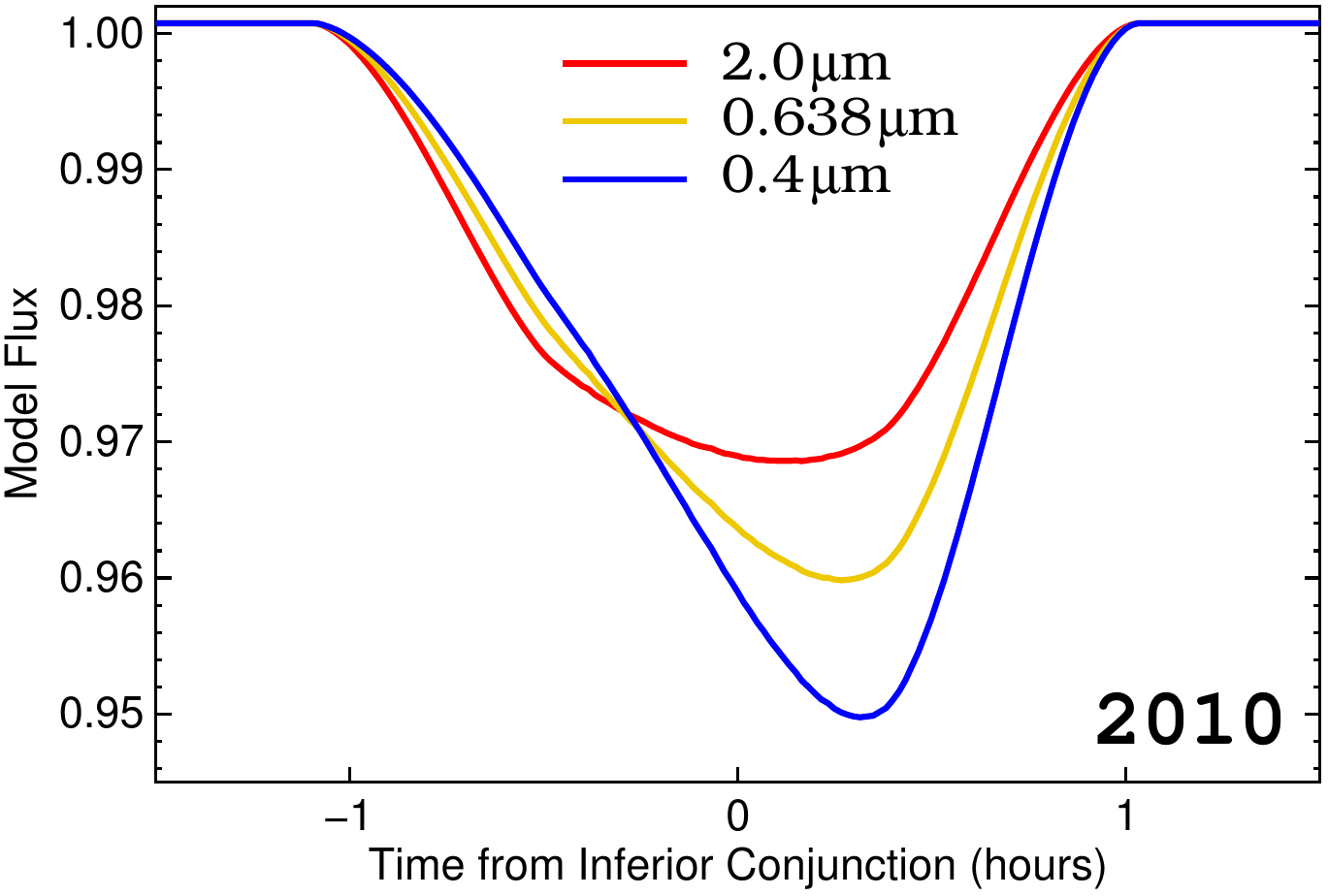}
\caption{\footnotesize Same as Figure \ref{figure:2009_colors}, but for 2010.
\label{figure:2010_colors}}
\end{figure}

\paragraph*{Chromatic Variation in Transit Shape}  Normally, transit lightcurve
shapes can show some variability with wavelength due to different degrees of
limb darkening.  But gravity darkening results from different effective
temperatures across the stellar disk.  Therefore transit shapes across
gravity-darkened stars show significant variation as a function of wavelength
\citep{2009ApJ...705..683B}.  We show predicted transit lightcurves for two
additional wavelengths ($0.4~\um$ and $2.0~\um$, in addition to the R-band
$0.638~\um$ from the \citet{2012ApJ...755...42V} photometry) in Figures
\ref{figure:2009_colors} and \ref{figure:2010_colors}.  Simultaneous multicolor
photometry could confirm the gravity darkening hypothesis and at the same time
help to constrain the stellar polar temperature and limb darkening parameters.

\section{DISCUSSION AND CONCLUSION}\label{section:discussion} 

We show that the unusual transit lightcurve shapes of \planet~and their
variation can be explained by a precessing planet transiting a gravity-darkened
star.  If our model is correct, then it would serve as a validation of
\planet~as a planet.  The \citet{2012ApJ...755...42V} discovery paper
established an upper limit on the mass of \planet~of $5.5~\mathrm{M_{Jup}}$, but
could not definitively rule out potential false positives.  No false-positive
scenario could reproduce the combination of gravity darkening and nodal
precession that we see in the \star~system.  We caution, though, that our
scenario would need to be verified using the effects from Section
\ref{section:projection} to fully confirm \planet~as a planet. \planet~stands to
be the second known Hot Jupiter orbiting an M-dwarf star after KOI-254b
\citep{2012AJ....143..111J}.  It would be the only transiting planet known to
orbit a T-Tauri star, and that star would be the youngest, coolest, and
lowest-mass star to host a transiting planet.

Our measured planetary radii, $1.64~\mathrm{R_{Jup}}$ and
$1.68~\mathrm{R_{Jup}}$, are smaller than that estimated by
\citet{2012ApJ...755...42V} ($1.91~\mathrm{R_{Jup}}$).  This difference owes
partially to our slightly smaller estimated stellar radius (1.03 or 1.04
$\mathrm{R_\odot}$ as opposed to $1.07~\mathrm{R_\odot}$), and partially to the
presumed higher fidelity of our gravity-darkened fit.

Our best-fit masses of $3.0~\mathrm{M_{Jup}}$ and $3.6~\mathrm{M_{Jup}}$ are
consistent with the \citet{2012ApJ...755...42V} radial-velocity-derived upper
limit of $5.5~\mathrm{M_{Jup}}$.  Interestingly, these masses and radii place
\planet~near to a Roche-lobe-limited state, just barely able to hold on to its
atmosphere against tidal disruption (as do the original values from 
\citet{2012ApJ...755...42V}, despite the different masses and radii).

The determination of both mass and radius for the planet allows us to place
constraints on its composition and thermal evolution.  At $3.0~\mathrm{M_{Jup}}$
and with $R_p=1.64~\mathrm{R_{Jup}}$, \planet~is clearly a hydrogen-dominated
gas giant, as it lies above the pure hydrogen curve from
\citet{2007ApJ...659.1661F}.  Furthermore, as a brand-new planet \planet~falls
very close to the predicted radius curve for a 10-Myr-old
$10~\mathrm{M_\oplus}$-core planet according to tables from
\citet{2007ApJ...659.1661F}\footnote{{\tt
http://www.ucolick.org/\~{ }jfortney/models.htm}}.  In particular, the high radius
of \planet~at such a young age rules out the ``cold start" model for compact
initial conditions outlined by \citet{2007ApJ...655..541M}.

Our fits show a large misalignment between the stellar spin and the planet orbit
of $69^\circ$.  Such a large misalignment may be inconsistent with the
assumption of synchronous stellar rotation.  Van Eyken et al. (2012) showed a
strong peak in a photometric periodogram corresponding to the planet's orbital
period of 0.448413 days.  Van Eyken et al. (2012) interpreted that peak to
represent the rotation period of the star, evident in the photometry due to
starspots.  Synchronous rotation predicts a stellar $v\sin i$ of 102~km/s with
our $M_*=0.34~\mathrm{M_\odot}$ best-fit values (similar to the
spectroscopically measured value of $80\pm8~\mathrm{km/s}$ from
\citet{2012ApJ...755...42V}).  Thus \star~is a fast-rotator.  However, truly
synchronous rotation might be difficult if not impossible to achieve via tidal
torques with $\spinorbit=69^\circ$.  Future photometry of transit shapes for
\planet~may allow for a dynamical fit for stellar rotation rate, which should
help to shed light on the accuracy of the synchronous stellar rotation
determination.

The high spin-orbit misalignment of $\spinorbit=69^\circ$ has implications for
planet formation and evolution.  \planet~is not the only near-polar-orbiting Hot
Jupiter \citep[\emph{e.g.}, Kepler-63b, ][]{PolarKeplerPlanet}.  But the
presence of such a highly inclined planet around such a young star supports the
idea from \citet{2010ApJ...718L.145W} that Hot Jupiters either form with random
orientations or very quickly acquire random orientations after their formation.
Furthermore, the young age of this highly inclined planet indicates that
\planet~could not have formed by Kozai resonance followed by tidal evolution, as
some Hot Jupiters may have \citep{2003ApJ...589..605W,2007ApJ...669.1298F}.  Any
planet-planet scattering event would necessarily also have been followed by some
degree of tidal evolution to circularize the orbit (\planet~cannot have an
eccentricity greater than $\sim0.5$, otherwise it would be entering the star on
each orbit), which would have been difficult given the available time.

One other extrasolar planet has been seen to be undergoing nodal precession: 
KOI-13b, as discovered by \citet{2012MNRAS.421L.122S}.  The algorithm that we
developed here might possibly be applied to KOI-13 and other systems like it in
the future.

\acknowledgements

\bibliographystyle{apj}
\bibliography{references}

\end{document}